\documentclass{aa}

\usepackage{graphicx}
\usepackage{graphicx}
\usepackage{subcaption}
\usepackage{txfonts}
\usepackage{natbib}

\usepackage{caption}
\usepackage{xcolor}     

\usepackage{rotating}
\usepackage{array} 

\newcolumntype{H}{>{\setbox0=\hbox\bgroup}c<{\egroup}@{}}

\DeclareCaptionLabelFormat{continued}{#1~#2 (Continued)}
\bibpunct{(}{)}{;}{a}{}{,}

\begin{document}

\title{Strongly lensed candidates from the HSC transient survey}

\author{Dani C.-Y. Chao\inst{1, 2, 3} \and James H.-H. Chan\inst{4} \and Sherry H. Suyu\inst{1,2,3} \and Naoki Yasuda\inst{5} \and Tomoki Morokuma\inst{6,5} \and Anton~T.~Jaelani\inst{7,8} \and Tohru Nagao\inst{9} \and C. E. Rusu\inst{10}}

\institute{Max Planck Institute for Astrophysics,
              Karl-Schwarzschild-Str. 1, 85741 Garching, Germany\\
              \email{dchao@mpa-garching.mpg.de}
              \and
              Physik-Department, Technische Universität München, 
              James-Franck-Straße 1, 85748 Garching, Germany
              \and
              Institute of Astronomy and Astrophysics, Academia Sinica, 
              11F of ASMAB, No.1, Section 4, Roosevelt Road, Taipei 10617, Taiwan
              \and
              Institute of Physics, Laboratory of Astrophysics, Ecole Polytechnique Fédérale de Lausanne (EPFL), 
              Observatoire de Sauverny, 1290, Versoix, Switzerland
              \and
              Kavli Institute for the Physics and Mathematics of the Universe (WPI), The University of Tokyo Institutes for Advanced Study, The University of Tokyo, 5-1-5 Kashiwanoha, Kashiwa,Chiba 277-8583, Japan
              \and
              Institute of Astronomy, Graduate School of Science, The University of Tokyo, 
              2-21-1 Osawa, Mitaka, Tokyo 181-0015, Japan
              \and
              Department of Physics, Kindai University, 
              3-4-1 Kowakae, Higashi-Osaka, Osaka 577-8502, Japan
              \and
              Astronomy Research Division and Bosscha Observatory, FMIPA, Institut Teknologi Bandung, 
              Jl. Ganesha 10, Bandung 40132, Indonesia
              \and
              Research Center for Space and Cosmic Evolution, Ehime University, 
              Matsuyama, Ehime 790-8577, Japan
              \and 
              National Astronomical Observatory of Japan, 2-21-1 Osawa, Mitaka, Tokyo 181-0015, Japan
              }

\date{Received --}

\abstract {We present a lensed quasar search based on the 
  variability of lens systems in the Hyper Suprime-Cam (HSC) transient survey. Starting
  from 101,353 variable objects with \textit{i}-band photometry in the
  HSC transient survey, we used a variability-based lens search
  method measuring the spatial extent in difference images to
  select potential lensed quasar candidates. We adopted conservative
  constraints in this variability selection
  and obtained 
  83,657 variable objects as possible lens candidates.
We then ran
  \textsc{Chitah}, a lens search algorithm based on the image
  configuration, on those 83,657 variable objects, and 2,130 variable
  objects were identified as potential lensed objects.  
  We visually
  inspected the 2,130 variable objects, and seven of them are our final
  lensed quasar candidates. Additionally, we found one lensed
  galaxy candidate as a serendipitous discovery. Among the eight final
  lensed candidates, one is the only known quadruply lensed quasar in the
  survey field, HSCJ095921+020638. None of the other
  seven lensed candidates have been previously classified as a lens nor a lensed candidate.
  Three of the five final candidates with available Hubble Space Telescope (HST) images, including HSCJ095921+020638, show 
  clues of a lensed feature in the HST images.    
  We show that a tightening of our variability selection criteria might result in the loss of possible 
  lensed quasar candidates, especially the lensed quasars with faint
  brightness or narrow separation, without efficiently eliminating the
  non-lensed objects; \textsc{Chitah} is therefore important as an advanced
  examination to improve the lens search efficiency through the object
  configuration. The recovery of HSCJ095921+020638 proves the
 effectiveness of the variability-based lens search method, and
  this lens search method can be used in other cadenced imaging
  surveys, such as the upcoming Rubin Observatory Legacy Survey of
  Space and Time.}
  

\maketitle
%

\section{Introduction}

Strong gravitational lensing is a useful tool to study the
Universe. Particularly, strong lensing of quasars can be used to
study various topics in astrophysics and cosmology, such as the
formation and evolution of supermassive black holes
\citep[e.g.][]{high_z_lensed_quasar}, quasar host properties
\citep[e.g.][]{quasar_host_01, quasar_host_02}, and the substructure
of dark matter \citep[e.g.][]{Gilman_Sub,anna_dm_sub}. Moreover,
combined with the time delays between lensed images, lensed quasars
can be used to measure the Hubble constant, $H_0$, which is crucial
for examining cosmological models and probing dark energy
\citep[e.g.][]{holicow_h0_01, holicow_h0_02}.

Enlarging the sample of lensed quasars is important for further
studies, and there have been several systematic searches looking for lensed
quasars. The Cosmic Lens All Sky Survey \citep[CLASS;][]{class_01,
  class_02} is the first systematic lensed quasar search in
radio wavelengths. 
CLASS identified lensed quasars by looking for resolved
multiple images in high-resolution radio images of sources that were preselected to have flat spectra.
In the optical, a large sample of lensed quasars has been constructed
in the SDSS\footnote{SDSS, Sloan Digital Sky Survey
  \citep{SDSS_survey}} Quasar Lens Search \citep[SQLS;][]{sqls_01,
  sqls_02, sqls_03, sqls_04, BQLS}. Starting with spectroscopic confirmed SDSS quasars, 
SQLS applied morphological and colour selections to find lens candidates. Nowadays, thanks to multiple 
large-scale surveys with a higher spatial resolution, larger covering area, and/or increased depth, more and more lensed
quasars have been found, and new lens search methods have been developed with the
aim to take advantage of those new surveys, such as catalogue
exploration techniques \citep[e.g.][]{Adriano_method,
  Adriano_alone, Strides_method, Peter_Williams_method, Rusu_18}, and inspections of the image configurations
\citep[e.g.][]{Adriano_method, chitah}. Citizen science is also
useful for lensed quasar searches \citep{space_warp_01, space_warp_02,
  ale_citizen}.

The \textit{Gaia}\footnote{\textit{Gaia}
  \citep{Gaia_survey}} all-sky survey provides new approaches for lensed quasar searches. 
In combination with other surveys, such as SDSS, PanSTARRS\footnote{Pan-STARRS, Panoramic Survey
  Telescope and Rapid Response System \citep{panstarrs_survey}}
\citep[e.g.][]{Cameron_01, Cameron_02, Cameron_03, Ostrovski_gaia_panstarrs_03},
DES\footnote{DES, Dark Energy Survey
  \citep{des_survey}} \citep[e.g.][]{Adriano_gaia_des_02,
  Adriano_gaia_des_01}, CRTS\footnote{CRTS, Catalina Real-Time Transient Survey \citep{CRTS}} \citep[e.g.][]{wise_gaia}, one could
search for lensed quasars by looking for \textit{Gaia} multiplets
or by comparing their flux and position offsets. One could also apply
machine learning techniques to look for lensed quasars within
\textit{Gaia} \citep{gaia_ml}.

Since quasars are variable sources, 
it is possible
to conduct lensed quasar searches through the exploitation of their
variability. A cadenced survey is ideally designed to reveal this variability through difference imaging. \citet{Kochanek_method} first proposed using
difference imaging to find lensed quasars:
while most of the (non-lensed) variable objects are point-like sources, lensed quasars with a sufficient angular size may appear as extended variable objects in the difference images. Specifically, lensed quasars whose image separations are comparable to the size of the point-spread function (PSF) appear as extended objects.  Small quasar-image separations (relative to the PSF) lead to point-like objects, whereas large quasar-image separations lead to multiple point-like objects in the difference image. Therefore, the objects that exhibit an extended feature or have multiple point features in the difference images are potential lenses. As a cadenced survey, the upcoming
Rubin Observatory Legacy Survey of Space and Time (LSST)\footnote{LSST \citep{lsst_survey}} will provide not only a huge pool of thousands of
new lensed quasars \citep[][hereafter OM10]{OM10}, but also a great opportunity to apply the time
variability for a lens search.

In this work, we conduct a lensed quasar search by applying the variability-based method described in \citet[][hereafter C20]{DC_Buddy} to the HSC\footnote{HSC, Hyper Suprime-Cam \citep{Miyazaki_hsc, Aihara_SSP}} transient survey \citep{hsc_transient}. The HSC transient survey is an ongoing
cadenced survey of the Subaru Telescope \citep{Miyazaki_subaru}, and
it has a similar image quality as expected for the LSST. The final
candidates of lensed quasars here undergo a three-step process: (1)
a variability-based selection (C20)
according to their spatial extent in the difference images of the HSC
transient survey; (2) classification as potential lenses by
\textsc{Chitah} \citep{chitah}, a lens search algorithm examining the
image configuration through lens modelling; and (3) visual inspection.
  
The organisation of this paper is as follows. In Sec.~\ref{sec:hsc},
we introduce the data from the HSC transient survey used in this
work. We describe our lens selection method in
Sec.~\ref{sec:method}. Sec.~\ref{sec:results} shows the lens selection
results, followed by our conclusions in Sec.~\ref{sec:conclusion}.

\section{The HSC transient survey}
\label{sec:hsc}

The HSC transient survey has observed the COSMOS \citep{COSMOS} field as part of the
HSC-SSP (Subaru Strategic Program; \citealt{Aihara_SSP,
  HSC_Miyazaki_2018,HSC_Komiyama_2018, HSC_Kawanomoto_2018,
  HSC_Furusawa_2018}) from November 2016 to June 2017, covering 1.77
$\text{deg}^2$ in the UltraDeep layer with a pixel size of 0.168$\arcsec$. 
There are 8, 9, 13, 14, and 11 epochs (i.e. nights of observations) in the \textit{g}-, \textit{r}-, \textit{i}-,
\textit{z}-, and \textit{y}- bands, respectively, with median depths of \textit{g}=26.4 mag,  \textit{r}=26.3 mag, \textit{i}=26.0 mag, \textit{z}=25.6 mag, and \textit{y}=24.6 mag. For the difference
imaging, the HSC transient survey uses the methods in
\citet{Robert_diff} and \citet{Alard_diff}.

Before we proceeded to the lens search with the difference images, we
first selected `HSC variables' from the HSC transient survey. An HSC
variable is defined in C20 as an object that is detected on
the difference images at least twice in the HSC transient survey -- the
detections could be from two different epochs or two different
bands. Following C20, we focussed on the HSC variables that have difference images 
in the \textit{i}-band, as the simulation and lens search algorithm have been developed 
for the \textit{i}-band in C20. From 2,252,293 sources contained in the COSMOS field \citep{cosmos_gal_2007, cosmos_source},
we picked out 101,353 sources with HSC \textit{i}-band difference images by requiring at least two detections in the HSC transient survey.

\section{Selection method}
\label{sec:method}
We selected lensed quasars in three steps. We first performed the
variability-based selection among 101,353 HSC variables
(Sec.~\ref{sec:hsc}) with their \textit{i}-band difference images,
which is described in Sec.~\ref{sec:time_variability}. After the
variability-based selection, we ran \textsc{Chitah} \citep{chitah}, a
lens search method based on the image configuration that is described
in Sec.~\ref{sec:chitah}. Finally, we visually inspected the
remaining HSC variables and graded them according to the scheme
described in Sec.~\ref{sec:grading}. The HSC variables with the highest
scores are our final lensed quasar candidates. In this work, we focus
 on quadruply lensed quasars (quad).

\subsection{Variability-based selection}
\label{sec:time_variability}
The variability-based lens search method described in C20 quantifies the
spatial extent of HSC variables on the \textit{i}-band
difference images, and it selects the HSC variables with a large spatial
extent as lensed quasar candidates. Briefly, the steps of
this method are as follows:
\begin{enumerate}
\item Create the `3$\sigma$-mask' for each HSC variable, $m$, in each epoch, $t$, by defining 
\begin{equation}
\label{eq:3sigma_mask}
I^m_{\text{mask}, t}(i, j) = 
\begin{cases}
1, & \text{if } |I^m_t(i, j)| > 3\sigma^m_t(i, j)\\
0, & \text{otherwise}
\end{cases} 
\end{equation}
where $i=1, ..., N_x$ and $j=1, ..., N_y$ are the pixel indices in the
difference image cutout $I^m_t(i,j)$ of dimensions $N_x \times
N_y$,\footnote{In this work, we use cutouts of $10\arcsec \times
  10\arcsec$ ($N_x=N_y=59$) in the variability-based selection.}
and $\sigma^m_t(i,j)$ are the estimated 1-$\sigma$ uncertainties in the
difference image cutout. Pixels $(i,j)$ with $I^m_{\text{mask}, t}(i,j)=1$ are the pixels in the 3$\sigma$-mask.
\item Define the `effective region', $I^m_{\text{eff}, t}(i, j)$, and
  the area of the effective region, $A^m_{\text{eff}, t}$, for each HSC
  variable $m$ in each epoch $t$ by
\begin{equation}
\label{eq:eff_mask}
I^m_{\text{eff}, t}(i, j) = 
\begin{cases}
1, & \text{if } \displaystyle\sum_{i'=i-1}^{i+1} \sum_{j'=j-1}^{j+1} I^m_{\text{mask},t}(i', j') > 2\\
0, & \text{otherwise}
\end{cases} 
\end{equation}
and
\begin{equation}
\label{eq:region_area}
A^m_{\text{eff}, t} =  \sum\limits_{i=1}^{N_x} \sum\limits_{j=1}^{N_y} I^m_{\text{eff}, t}(i, j).
\end{equation}
where $A^m_{\text{eff}, t}$ is equivalent to the number of pixels with
pixel values equal to one in the effective region ($I^m_{\text{eff},
  t}=1$). While the 3$\sigma$-mask includes many isolated single pixels which are actually noise peaks,
  the effective region area, $A^m_{\text{eff}, t}$, is a more accurate quantification of the object's spatial extent in the difference image.\footnote{The RA, DEC, and effective region areas of the 101,353 HSC variables investigated in this paper are available at \url{https://github.com/danichao/HSC_Variable}.} 
\item Compute $A_{\text{eff},t}(p_\text{thrs}\%)$ as the effective region area for each epoch $t$ such that
$p_\text{thrs}\%$ of the HSC variables have values $A^m_{\text{eff},t} < A_{\text{eff},t}(p_\text{thrs}\%)$. 
If an HSC variable $m'$ is not observed in an epoch $t'$, $A^{m'}_{\text{eff},t'}=0$ would be assigned. 
For such case, we did not include it when we computed $A_{\text{eff},t}(p_\text{thrs}\%)$.
\item Set a threshold on the number of epochs, $N_\text{thrs}$, such that an HSC variable $m$ is selected as a lensed quasar candidate if it satisfies
\begin{equation}
\label{eq:n_thrs}
A^m_{\text{eff},t} > A_{\text{eff},t}(p_\text{thrs}\%)
\end{equation}
for more than $N_\text{thrs}$ epochs.
Table~\ref{tab:HSC_epochs} shows the dates and seeings of all the 13 epochs in the \textit{i}-band.
\end{enumerate}
For the first application of C20, we employed very loose constraints,
$p_\text{thrs}\%=50\%$ and $N_\text{thrs}=0$, to select lensed quasar
candidates. Among the 101,353 HSC variables, 83,657 HSC
variables were identified as lensed quasar candidates by the
variability-based selection. The variability-based selection did not bring down the number of potential lensed quasar candidates substantially due to these loose constraints. However, the variability-based selection is important for finding lensed quasars. Lensed quasars are not only variable, but also multiple point-like or extended if not deblended. These loose constraints already discarded part of the contaminations, such as some single point-like variables and false detections, and shall become more stringent if we attempt to find lensed quasars with the variability-based lens search algorithm in a field that is much larger than COSMOS. By properly tightening the constraints in the variability-based selection, we will be able to improve the lens search efficiency (see the discussion in Sec.~\ref{sec:discussion}).

\begin{table}
\caption{HSC transient survey observation dates and seeings in the \textit{i}-band.}
\label{tab:HSC_epochs}
\centering
\begin{tabular}{ c c }
\hline
Epochs/Observation dates & Seeing (arcsec) \\ 
\hline          
2016-11-25 & 0.83 \\
2016-11-29 & 1.16\\
2016-12-25 & 1.25\\
2017-01-02 & 0.68\\
2017-01-23 & 0.70\\
2017-01-30 & 0.76\\
2017-02-02 & 0.48\\
2017-02-25 & 0.72\\
2017-03-04 & 0.69\\
2017-03-23 & 0.66\\
2017-03-30 & 0.98\\
2017-04-26 & 1.24\\
2017-04-27 & 0.58\\            
\hline
\end{tabular}
\end{table}

\subsection{\textsc{CHITAH}}
\label{sec:chitah}
For the 83,657 lensed quasar candidates selected in
Sec.~\ref{sec:time_variability}, we ran \textsc{Chitah} on their
stacked images from the HSC survey (\textit{g}-, \textit{r}-,
\textit{i}-, \textit{z}-, and \textit{y}-bands) to reduce the number
of lensed quasar candidates. Briefly, \textsc{Chitah} works as
follows: (1) picking two image cutouts, one from the bluer bands
  (\textit{g}/\textit{r}) and one from the redder bands
  (\textit{z}/\textit{y}) which have sharper PSFs; 
  (2) matching PSFs in the two selected bands;
  (3) decomposing the image cutouts into $P$ and $Q$, the former for the lens galaxy and 
the latter for the lensed images, according to colour information;
(4) estimating the lens photo-centre using 
the light distribution on $P$ and identify the lensed image positions on
$Q$ with four PSFs; and
(5) using the lensed image positions identified on $Q$ to model the lens mass distribution
with a singular isothermal ellipsoid (SIE) model \citep{SIE}.
%
%
%
%
%
%
%
%
%
%

The outputs of the model are the best-fitting SIE parameters: the
Einstein radius ($\theta_\text{Ein}$), the axis ratio ($q$), the
position angle (PA), and the centre of mass (of the lens). The convergence $\kappa$ of
the SIE model is given by
\begin{equation}
\label{eq:SIE}
\kappa (\theta_1, \theta_2) = \frac{\theta_\text{Ein}}{2\sqrt{\theta^2_1+\theta^2_2/q^2}},
\end{equation}
where $(\theta_1, \theta_2)$ are the coordinates relative to the centre of mass
along the semi-major and semi-minor axes of the elliptical mass distribution.
We determineed the SIE parameters by
minimising the $\chi^2_\text{source}$ on the source plane, which is
defined as
\begin{equation}
\label{eq:chi_source}
\chi^2_\text{source} = \sum \limits_n \frac{|\mathbf{r}_n-\mathbf{r}_\text{model}|^2}{\sigma^2_\text{image}/\mu_n},
\end{equation}
where $\mathbf{r}_n$ is the respective source position mapped from the
position of lensed image $n$ by the SIE lens model, $\mu_n$ is the
magnification of lensed image $n$ from the SIE lens model,
$\sigma_\text{image}$ was chosen to be the HSC pixel size
(0.168\arcsec) as an estimate of the uncertainty, and
$\mathbf{r}_\text{model}$ is the modelled source position that is
evaluated by a weighted mean of $\mathbf{r}_n$
\citep{Masamune_glafic},
\begin{equation}
\label{eq:r_model}
\mathbf{r}_\text{model}=\frac{\sum \limits_n \sqrt{\mu_n}\mathbf{r}_n}{\sum \limits_n \sqrt{\mu_n}}.
\end{equation}
Here the index $n$ runs from one to four for quads in this work. We also used the
lens photo-centre, $\mathbf{x}_\text{centre}$, from the light distribution on $P$
as a prior to constrain the centre of mass
of the SIE model, $\mathbf{x}_\text{model}$ \footnote{Previous studies \citep[e.g.][]{Koopman_centre} have shown that 
the offset between the photo centre and the centre of mass of isolated lenses is small, $\lesssim$0.05$\arcsec$.}. Therefore, we define
\begin{equation}
\label{eq:chi_centre}
\chi^2_\text{centre}=\frac{|\mathbf{x}_\text{model}-\mathbf{x}_\text{centre}|^2}{\sigma^2_\text{centre}},
\end{equation}
where $\sigma_\text{centre}$ was chosen to be the same 
as $\sigma_\text{image}$. We further took into account the residuals of
the fit to the lensed quasar image from \textsc{Chitah}. The difference
between the lensed image, $Q(i, j)$, and the predicted image formed by four PSFs,
$Q^\text{P}(i, j)$, is defined as
\begin{equation}
\label{eq:chi_res}
\chi^2_\text{residual}=\sum \limits_{i,j} \frac{\left [ Q(i, j) - Q^\text{P}(i, j) \right ]^2}{var (i, j)},
\end{equation}
where $i= 1, ..., N_x$ and $j= 1, ..., N_y$ are the pixel indices in
the image cutout of dimensions $N_x \times
N_y$,\footnote{In this work, \textsc{Chitah} uses cutouts of $7\arcsec \times
  7\arcsec (N_x=N_y=43)$.} and $var (i, j)$ is the pixel uncertainty in $Q(i,
j)$. In this work, we assume that $var (i, j)$ is constant and thus 
irrelevant in the minimisation for the point source positions. We note that
$Q^\text{P}$ was obtained from the four PSFs fitting in order to identify the lensed image positions,
which is independent of lens modelling. Therefore, the fitted fluxes of lensed images allow for the presence of image flux anomalies.

The criteria for the classification of lensed quasar candidates are 
\begin{equation}
\label{eq:cri_1}
\chi^2=\chi^2_\text{source} + \chi^2_\text{centre} < 1
\end{equation}
and
\begin{equation}
\label{eq:cri_2}
\chi^2_\text{residual} < 2.
\end{equation}
These two criteria allow \textsc{CHITAH} to extract a manageable number of 
candidates and a low false-positive rate ($< 3\%$, see Figure 1 in \citet{chitah}). We note that
the second criterion (Eq.~\ref{eq:cri_2}) is chosen empirically due to
the arbitrary scale in the pixel uncertainty. Consequently, this loose
constraint discarded only $\approx 2\%$ of the objects while most of the objects
($\approx 95\%$) were eliminated by Eq.~\ref{eq:cri_1}. The lensed 
candidates were selected by $0.3\arcsec < \theta_\text{Ein} < 2\arcsec$ and $q > 0.2$.  
After running
\textsc{Chitah}, we reduced the number of the lensed quasar candidates
from 83,657 to 2,130.

\subsection{Visual inspection}
\label{sec:grading}
We visually inspected the remaining 2,130 lensed quasar candidates. 
We first discarded 2,065 lensed quasar candidates that are obviously not lenses. 
Three of the coauthors then independently graded each of the 65 remaining lensed quasar candidates with 
the following grading scheme:

\begin{itemize}
\item 3: definite lens,
\item 2: probable lens,
\item 1: likely lens, and
\item 0: not a lens.
\end{itemize} 
The visual inspection was mainly conducted with the HSC colour-composite images. 
Typical aspects taken into consideration in grading are the colour
difference between the possible lensed images and the possible lens
galaxy, and the positions of the possible lensed images. This grading
scheme is the same as the one 
on the SuGOHI lens sample from the HSC survey \citep[e.g.][]{sugohi, ken_sugohi, chitah_hsc, anton_hsc}.
We then took the average score among the three
graders for each lensed quasar candidate, and the candidates with
average scores higher than 1.5 would be our final lensed quasar
candidates. We note that in visual inspection, the gradings
are based on general lensed features, which are not specific to lensed quasars. In total,
we have eight candidates with an average score higher than 1.5 (listed in Table~\ref{tab:candidate}). 
Further properties of the eight final candidates are listed in Table~\ref{tab:candidate_selection}.
One of our eight final candidates was found by chance (See Sec.~\ref{sec:candidates}). 
%

%

\begin{table*}
\caption{Candidates with average scores higher than 1.5 in visual inspection.}
\label{tab:candidate}
\centering
\begin{tabular}{ccccc}
\hline
Name             & RA (deg)         & DEC (deg)       &Average score   &   Comment  \\
\hline
HSCJ095921+020638  &    149.84071 & 2.11068 &    3.0                 &      \citet{Anguita_lens}     \\
HSCJ100050+013251  &    150.20947 & 1.54775 &    2.3            &      Probable lensed galaxy  \\
HSCJ095921+025700  &    149.84037 & 2.95013 &    2.0                &     Probable lensed galaxy. No HST image  \\
HSCJ100307+020241  &    150.78256 & 2.04484 &    1.7             &   No HST image        \\
HSCJ095943+022046  &    149.93252 & 2.34623 &   1.7             &      \\
HSCJ095744+023835  &    149.43571 & 2.64332 &    1.7             &          \\
HSCJ100050+031825  &    150.21200 & 3.30706 &   1.7              &    No HST image   \\
HSCJ100129+024427  &    150.37457 & 2.74093 &    2.0              &     Lensed galaxy candidate by serendipitous discovery   \\                  
\hline
\end{tabular}
\tablefoot{The table lists: Name, RA, DEC, and average score from visual inspection. HSCJ095921+020638 is the only known quadruply lensed quasar in the COSMOS field. HSCJ100050+013251, HSCJ095921+025700, and HSCJ100129+024427 are more likely to be lensed galaxy candidates. Only five of these objects have HST images (see Fig.~\ref{fig:hst_imgs} and \ref{fig:hst_wfc3}).}
\end{table*}


\begin{table*}
\caption{Variability selection and \textsc{Chitah} properties of the candidates listed in Table~\ref{tab:candidate}.}
\label{tab:candidate_selection}
\centering
\begin{tabular}{ccccccc}
\hline
Name               & Number of              & $\chi^2_\text{residual}$ & $\chi^2$  &  $\theta_\text{Ein}$  &   q   &    $\theta^\text{s}_\text{Ein}$         \\
                        & variability epochs &           &                                   &          [$\arcsec$]      &       &          [\arcsec]              \\
\hline
HSCJ095921+020638    & 7   & 0.000357   & 0.4    &   0.7  &  0.95 &  0.7\\
HSCJ100050+013251  &  6   & 0.013854  & 0.7    &   1.1  &  0.90   &  1.0 \\
HSCJ095921+025700  & 1   & 0.005372  & 1.0     &   1.2  &   0.64  &   0.9 \\
HSCJ100307+020241  & 7   & 0.001768  & 1.0    &   1.1   &   0.57  &   0.8  \\
HSCJ095943+022046  &  8   & 0.000225  & 0.3   &   0.9   &   0.75  &   0.8 \\
HSCJ095744+023835   & 5   & 0.000332   & 0.6  &   1.1   &   0.88  &   1.0  \\
HSCJ100050+031825  &  7   & 0.002117  & 0.3    &   1.0   &   0.89 &    0.9 \\
HSCJ100129+024427  & 0   & 0.003324   & 0.7    &   0.9    &    0.96 &  0.9 \\                  
\hline
\end{tabular}
\tablefoot{The table lists: Number of epochs where the final candidates satisfy the variability selection criterion, and their values of $\chi_{\rm residual}^2$, $\chi^2$, Einstein radius ($\theta_\text{Ein}$), axis ratio ($q$), and scaled Einstein radius ($\theta^\text{s}_\text{Ein}$) from \textsc{Chitah}. The scaled Einstein radius, $\theta^\text{s}_\text{Ein}$, is defined as $\theta^\text{s}_\text{Ein} = \theta_\text{Ein}\sqrt{2q^2/(1+q^2)}$.}
\end{table*}

\section{Results and discussion}
\label{sec:results}
In this section, we take a further look at the HSC variables that were selected by 
our approach. 
We describe the final candidates in
Sec.~\ref{sec:candidates}. In Sec.~\ref{sec:var_fp}, we discuss the
objects selected as candidates by both the variability selection and
\textsc{Chitah}, but rejected by our visual inspection. The discussion
about the objects that meet the variability selection criteria but
get rejected by \textsc{Chitah} is in Sec.~\ref{sec:chitah_fp}.We compare our lens candidates to previously identified candidates from other searches in Sec.~\ref{sec:discussion}.

\subsection{Final candidates}
\label{sec:candidates}
We show the eight final candidates in Fig.~\ref{fig:final_candidates} and Table~\ref{tab:candidate}.\footnote{If a variable meets the selection criteria at multiple epochs, we show the epoch with seeing that is closest to the median seeing ($0.72\arcsec$) as C20 shows that the variability selection has better lens search performance at the epochs with seeings close to the median seeing.}
The highest scored candidate, HSCJ095921+020638, is a known lensed
quasar \citep{Anguita_lens}. The recovery of HSCJ095921+020638
demonstrates the effectiveness of our variability-based lens searching
method. 

HSCJ095921+025700 was picked out by the variability selection
due to the loose criteria ($p_\text{thrs}\%=50\%$ and
$N_\text{thrs}=0$). Under these loose criteria, an HSC variable can
be selected as long as it has a non-zero effective region in one of
a few specific epochs, even if the effective region comes from the
accumulation of noise peaks. These objects that were selected due to possible noise peaks and further identified as lens candidates by their image configuration in colour-composite images are more likely to be lensed galaxy candidates, instead of lensed quasar candidates. Although the loose criteria are sensitive
to noise peaks, they are still sufficient for the COSMOS field we
examine in this work, given the covering area of the COSMOS field. Moreover, the loose criteria allow us to have a
more complete sample that includes lensed quasars with a faint
brightness or small separation.

Most of the other final candidates have substantial effective region areas satisfying the 
loose criteria in the variability selection, and also show a possible lens feature such as a colour gradient in the colour-composite images.
Thanks to a fortunate mistake, our variability-based method picked up HSCJ100129+024427 in an earlier version of our code.
HSCJ100129+024427 has a zero effective
region area across all the 13 epochs, and 
the final variability selection actually did not pick it out.
Given its lens-like appearance, we keep this system for further
investigation:
\textsc{Chitah} identifies the lensing feature
of HSCJ100129+024427, followed by an average score of probable lens
from the visual inspection. Therefore, HSCJ100129+024427 is more
likely a candidate of a lensed galaxy, instead of a candidate of a lensed
quasar.

We show archival Hubble Space Telescope (HST) F814W images of
HSCJ095921+020638, HSCJ100129+024427, HSCJ095943+022046, and
HSCJ095744+023835 (HST proposal id: 9822; PI: Scoville) in Fig.~\ref{fig:hst_imgs} \citep{hst_treasure}, and the HST F105W image of HSCJ100050+013251 (HST proposal id: 14808; PI: Suzuki) in Fig.~\ref{fig:hst_wfc3}. From the HST images, we
can clearly see the four multiple images in HSCJ095921+020638
(top-left panel in Fig.~\ref{fig:hst_imgs}), and three possible lensed
images in HSCJ100129+024427 (top-right panel in
Fig.~\ref{fig:hst_imgs}). We also see the possible arc feature in HSCJ100050+013251 (Fig.~\ref{fig:hst_wfc3}), indicating that this system is likely to be a lensed galaxy.   
On the other hand, the lensed features of
HSCJ095943+022046 and HSCJ100050+031825 are hard to see, and these two
objects are more likely to be spiral galaxies or galaxies with dust
lanes.
\begin{figure*}
\centering
\includegraphics[scale=0.3]{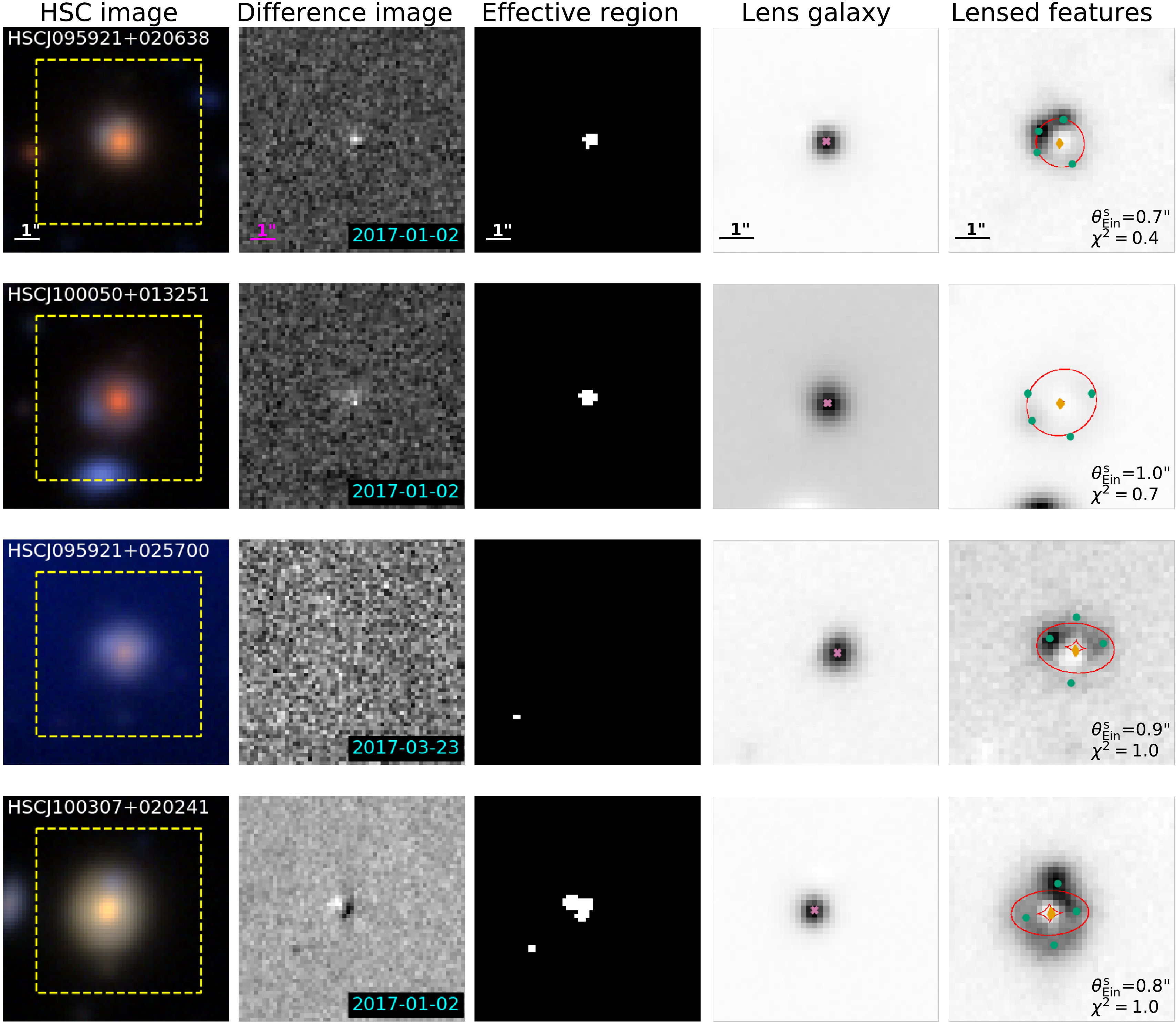}
\caption[]{Final candidates, the variable objects identified as lensed
  quasars by both the variability-based method and
  \textsc{Chitah} with average scores higher than 1.5 from the visual
  inspection. The first column shows the \textit{gri} image of each
  final candidate. The second column shows their difference images at
  the epoch at which they met the time variability selection criteria,
  and the third column shows their corresponding effective region, a
  quantification of the object's spatial extent in the
  difference image. The fourth column shows the lens galaxy predicted
  by \textsc{Chitah} with the lens centre  
  marked by the magenta cross. The fifth column shows their predicted lensed
  features by \textsc{Chitah} with their best-fit SIE model.
  The positions of the fitted lensed images are in green dots, the 
  predicted source position is labelled as the yellow diamond, and the
  red curves are the critical and caustic curves of their best-fit
  SIE model. The variability-based selection uses cutouts of
  10$\arcsec$ $\times$ 10$\arcsec$ (first three columns), and \textsc{Chitah} uses cutouts
  of 7$\arcsec$ $\times$ 7$\arcsec$, which is the yellow dashed box in
  the first column.} 
\label{fig:final_candidates}
\end{figure*}
\begin{figure*}
\ContinuedFloat
\centering
\includegraphics[scale=0.3]{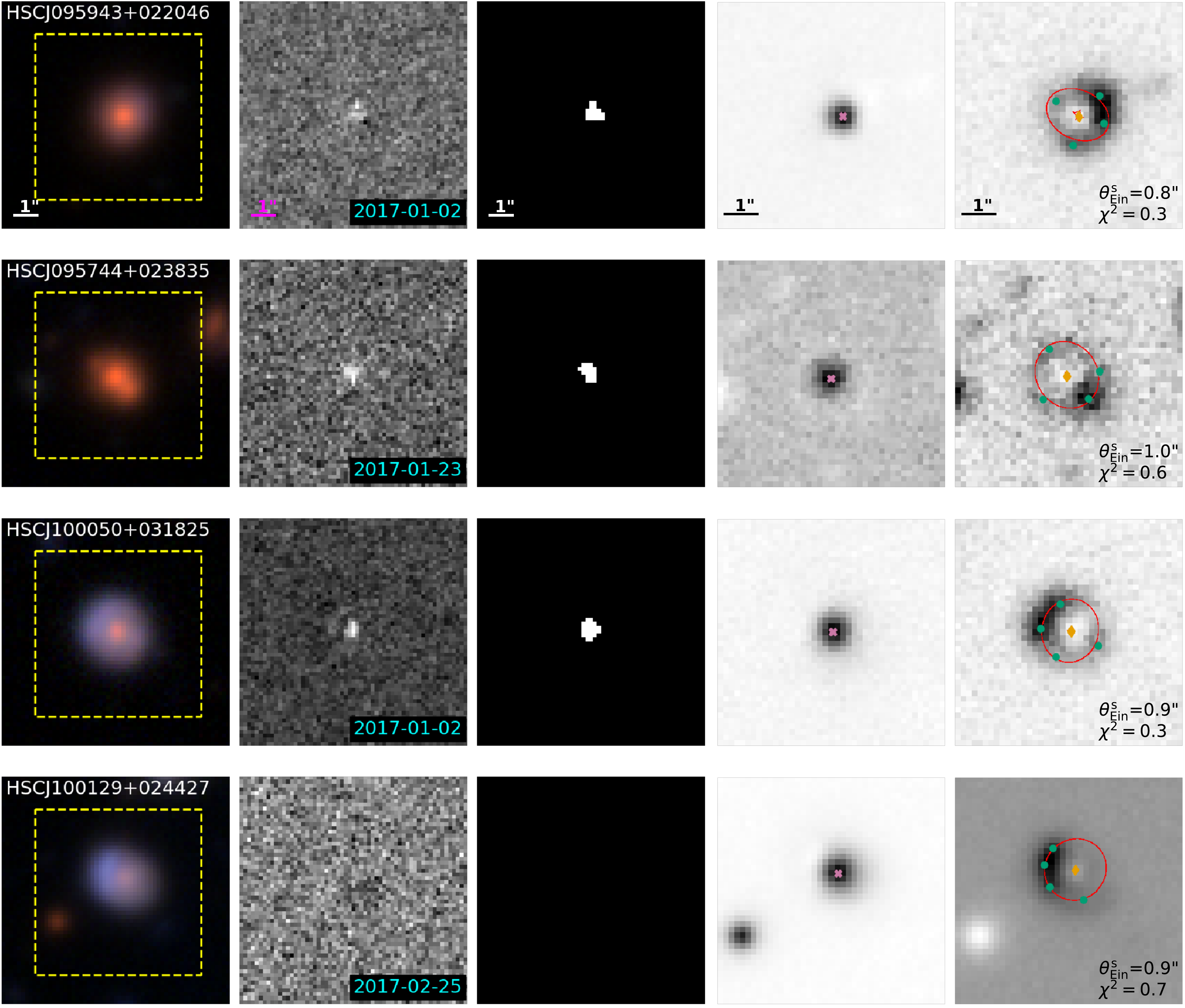}
\caption{HSCJ100129+024427 is a serendipitous discovery and is more likely to be a
candidate of a lensed galaxy due to the zero effective region area across all 13 epochs. See Sec.~\ref{sec:candidates} for details.}
\label{fig:final_candidates_con}
\end{figure*}

%
\begin{figure}
\includegraphics[width=\columnwidth]{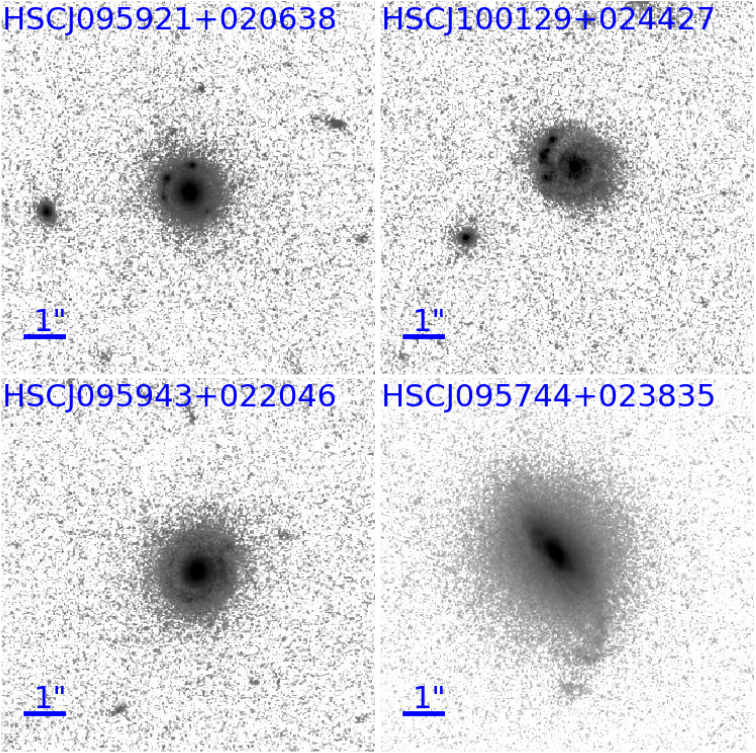}
\caption{HST F814W images of four final
  candidates. HSCJ095921+020638 (top left) is the only known quadruply
  lensed quasar in the COSMOS field.
}
\label{fig:hst_imgs}
\end{figure}

\begin{figure}
\includegraphics[scale=0.65]{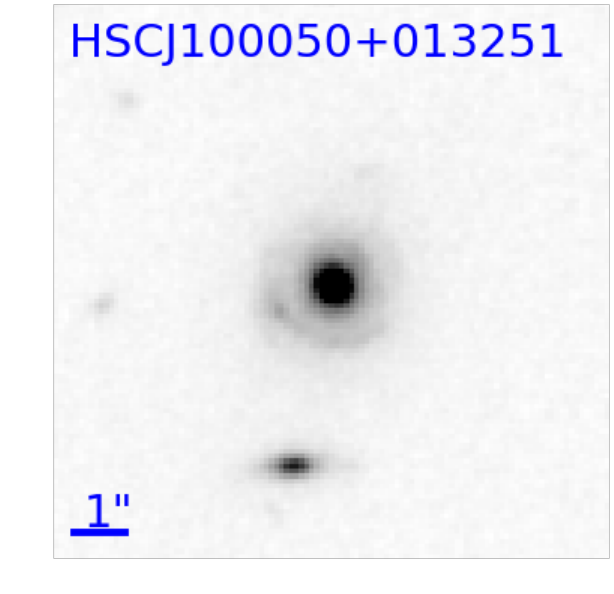}
\caption{HST F105W image of HSCJ100050+013251, a possible lensed galaxy. The possible arc is clearly visible.}
\label{fig:hst_wfc3}
\end{figure}

\subsection{\textsc{Chitah} false positives}
\label{sec:var_fp}
\begin{figure*}
\centering
\includegraphics[scale=0.285]{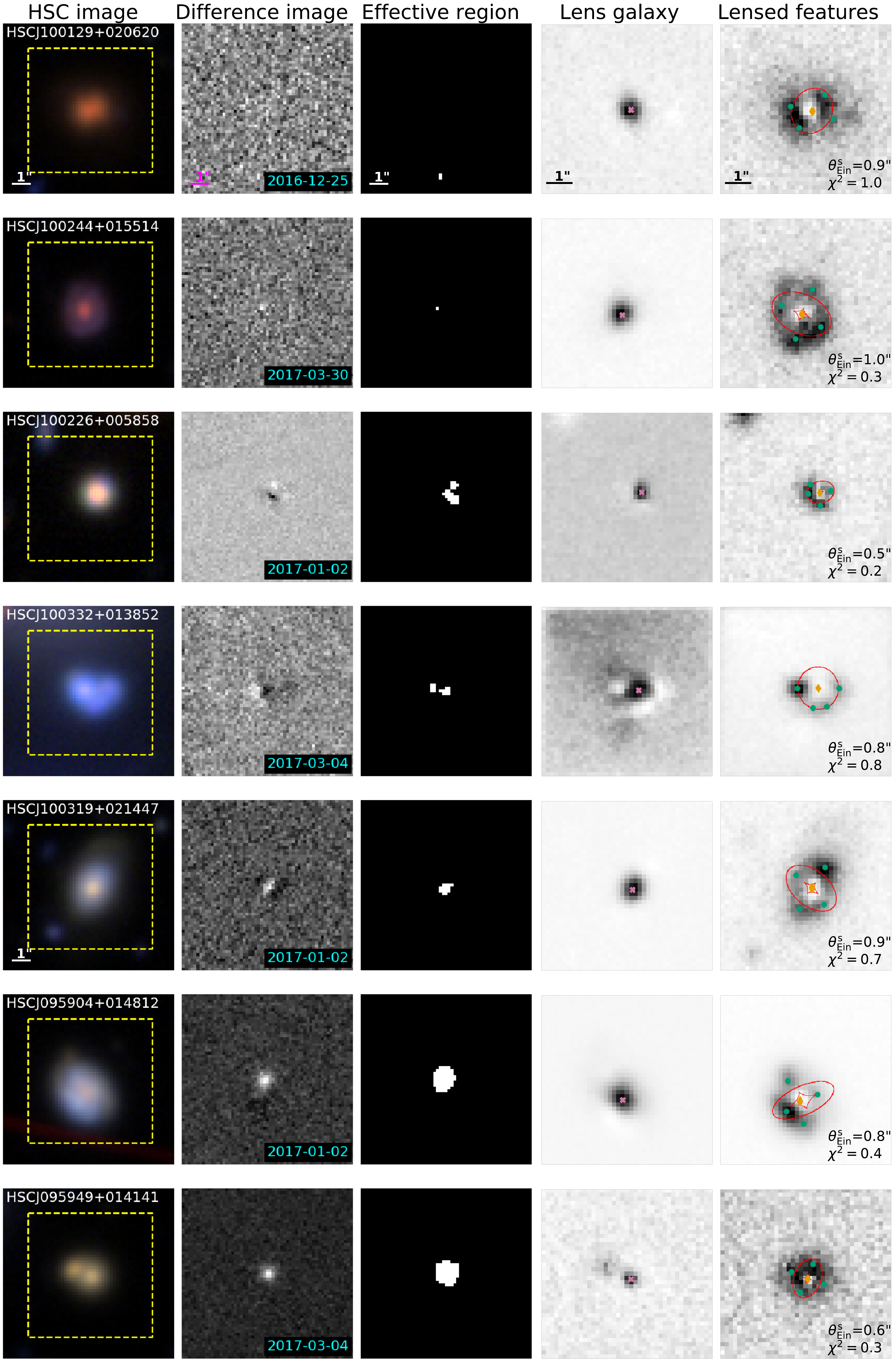}
\caption{Examples of \textsc{Chitah} false positives, the variable
  objects identified as lensed quasars by both the
  variability-based method and \textsc{Chitah}, but rejected by
  the visual inspection. The five columns are in the same format as in Fig.~\ref{fig:final_candidates}.
  For details about these 
 \textsc{Chitah} false positives, see Sec.~\ref{sec:var_fp}.
  }
\label{fig:visual_fp}
\end{figure*}

In Fig.~\ref{fig:visual_fp}, we show examples of the objects that get selected by both the variability-based selection and \textsc{Chitah},
but rejected in the visual inspection. Those
objects are called \textsc{Chitah} false positives in this
work.

These objects are classified as non-lenses based on visual inspection as 
they do not exhibit typical morphologies of lenses. For example, the system HSCJ100129+020620 
does not show lens-like arcs in the colour images, and the system HSCJ100244+015514 is probably a ring galaxy 
(with the ring being physically associated with the central galaxy) since the morphology of the ring is unlikely to be formed by lensing. Actually,
both HSCJ100129+020620 and HSCJ100244+015514 at only one epoch have
effective regions larger than
$A_{\text{eff},t}(p_\text{thrs}\%=50\%)$, and their effective regions
are only 1-2 pixels at those epochs (December 25, 2016 and March 30, 2017,
respectively). Moreover, the effective region of HSCJ100129+020620
more likely comes from noise peaks. Although \textsc{Chitah} might misidentify
kinds of objects such as HSCJ100129+020620 and HSCJ100244+015514, applying
stricter criteria in the variability-based selection (higher values for $p_\text{thrs}\%$ or $N_\text{thrs}$)
can avoid such misidentification. However, we might lose the candidates of faint or small-separation 
lensed quasars by doing so.

Compared to the final candidates and to the other \textsc{Chitah}
false positives, HSCJ100226+005858 exhibits a different colour in the
\textit{gri}-composite image, and it has a rather round and compact
shape. In fact, HSCJ100226+005858 is classified as a star in the
internal HSC transient catalogue
\citep{hsc_transient}. HSCJ100226+005858 is probably a variable star
or a star that is too bright for the transient pipeline to subtract
the light perfectly since it has substantial amounts of the effective
regions across almost all eight epochs at which it has been observed
in the \textit{i}-band (see Sec.~\ref{sec:chitah_fp} for detail).

The nearly monotone colour of HSCJ100332+013852 indicates that it is
unlikely to be a lensed object. From the \textit{gri}-composite image,
HSCJ100332+013852 is likely a merger of two or more galaxies. Such
a merger process might trigger AGN activity that results in effective
regions passing our selection criteria.

HSCJ100319+021447 shows a possible tidal feature around the top right,
and the possible tidal event could be the reason for the variable brightness
given previous studies that indicate mergers could trigger AGN activity and show tidal features
(e.g. \citealt{ellison_merger}, \citealt{silverman_merger}, \citealt{capelo_merger}, \citealt{commerford_merger}, \citealt{stemo_merger}).
 Moreover, \textsc{Chitah} identifies the blue trait around
this possible tidal feature as the lensed images in HSCJ100319+021447,
hence the misidentification of HSCJ100319+021447. 

HSCJ095904+014812 is
classified as a supernova in the internal HSC transient catalogue, and
it exploded during the survey period in Table~\ref{tab:HSC_epochs},
resulting in the substantial effective region areas. Furthermore, the
blue features in HSCJ095904+014812 are likely star-forming regions of
a spiral galaxy, and 
these blue features cause the misidentification from
\textsc{Chitah}. HSCJ095949+014141 is also classified as a supernova
in the internal HSC transient catalogue. While HSCJ095949+014141 appears
to be binary, \textsc{Chitah} misidentifies it as a lens because of the
colour gradient in the bottom-right object (so part of the object is
mistaken as lensed features) or imperfect PSF matching. 
Additionally, when checking the difference images and 
the effective regions of both HSCJ095904+014812 and HSCJ095949+014141,
we found three stages: (1) before the exploding epochs, nothing was visible 
in the difference images and the corresponding 
effective region areas were also zero; (2) the effective region areas
increased suddenly and significantly in an epoch, and continued to increase afterwards;
(3) after the effective region areas reached a maximum in an epoch,
the effective region areas started to decrease.  
These three stages are similar to a supernova,
indicating that the effective region (Eq.~\ref{eq:eff_mask}) can also be used for the detection of supernovae.
The combination of the variability-based selection and \textsc{Chitah} is even possible 
for the detection of lensed supernovae, though the detection might happen at a much later time after the explosion.

\subsection{Variability false positives}
\label{sec:chitah_fp}
\begin{figure}
\centering
\includegraphics[width=\columnwidth]{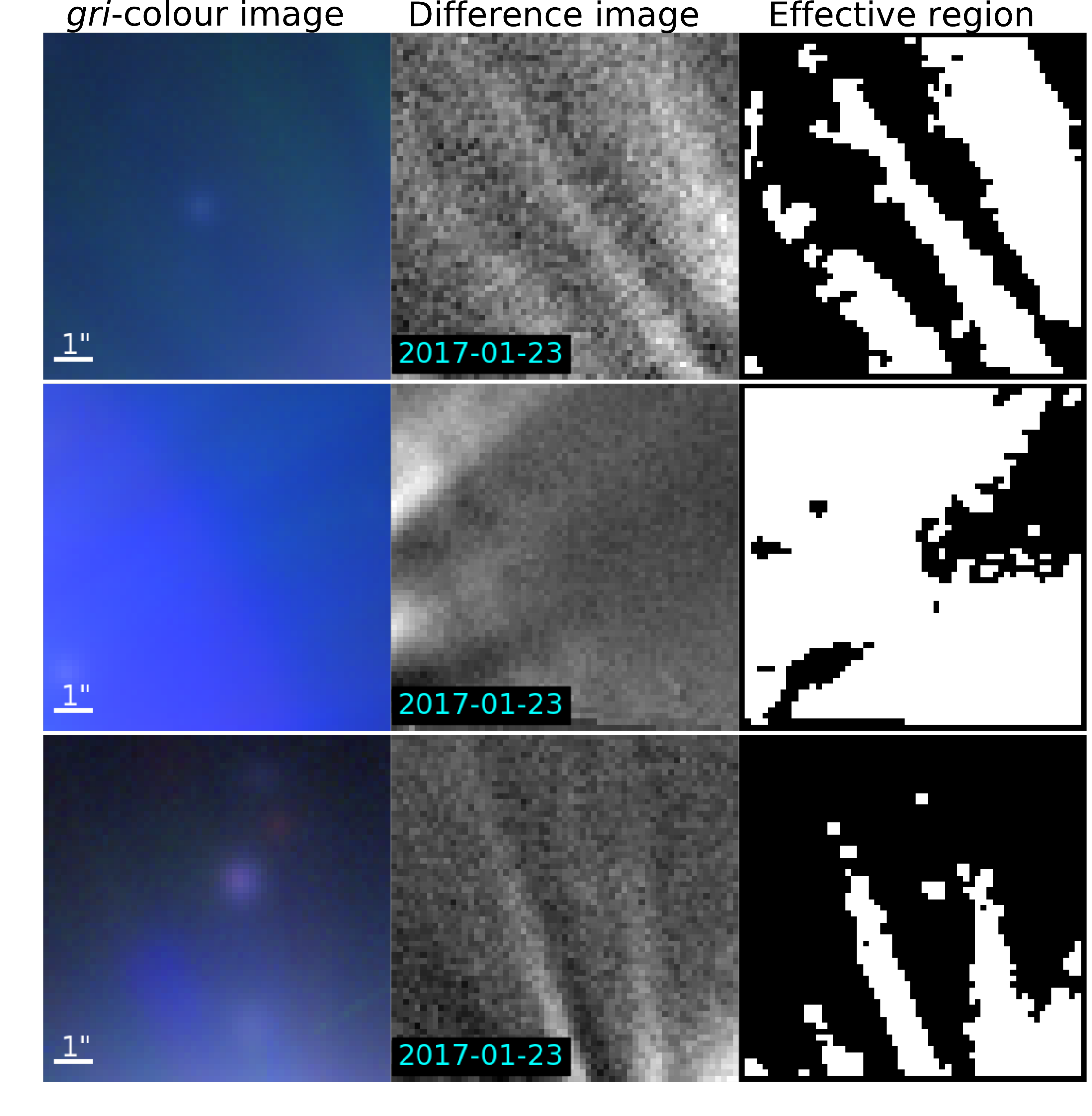}
\caption{Artificial effects that were selected by the
  variability-based selection due to the extremely large
  effective regions.}
\label{fig:CA}
\end{figure}

Here we examine the objects selected by the variability-based
selection but rejected by \textsc{Chitah}, and we call these objects
variability false positives in this work. Many of variability
false positives have extremely large effective region areas across all
the epochs, which are due to artificial effects, as the
examples shown in Fig.~\ref{fig:CA}. A misidentification such as the one in
Fig.~\ref{fig:CA} could be prevented by assigning an upper limit to
the effective region area. We can discard an HSC variable from the
lens candidate selection if its effective region area exceeds the
upper limit in a certain number of epochs since an abnormally large area of effective regions 
over many epochs suggests that heavy artificial effects
happen around the HSC variable and that the detection of the HSC
variable is not reliable. By doing so, however, we might miss lens candidates that are located close to artefacts. 

\begin{figure*}
\centering
\includegraphics[scale=0.3]{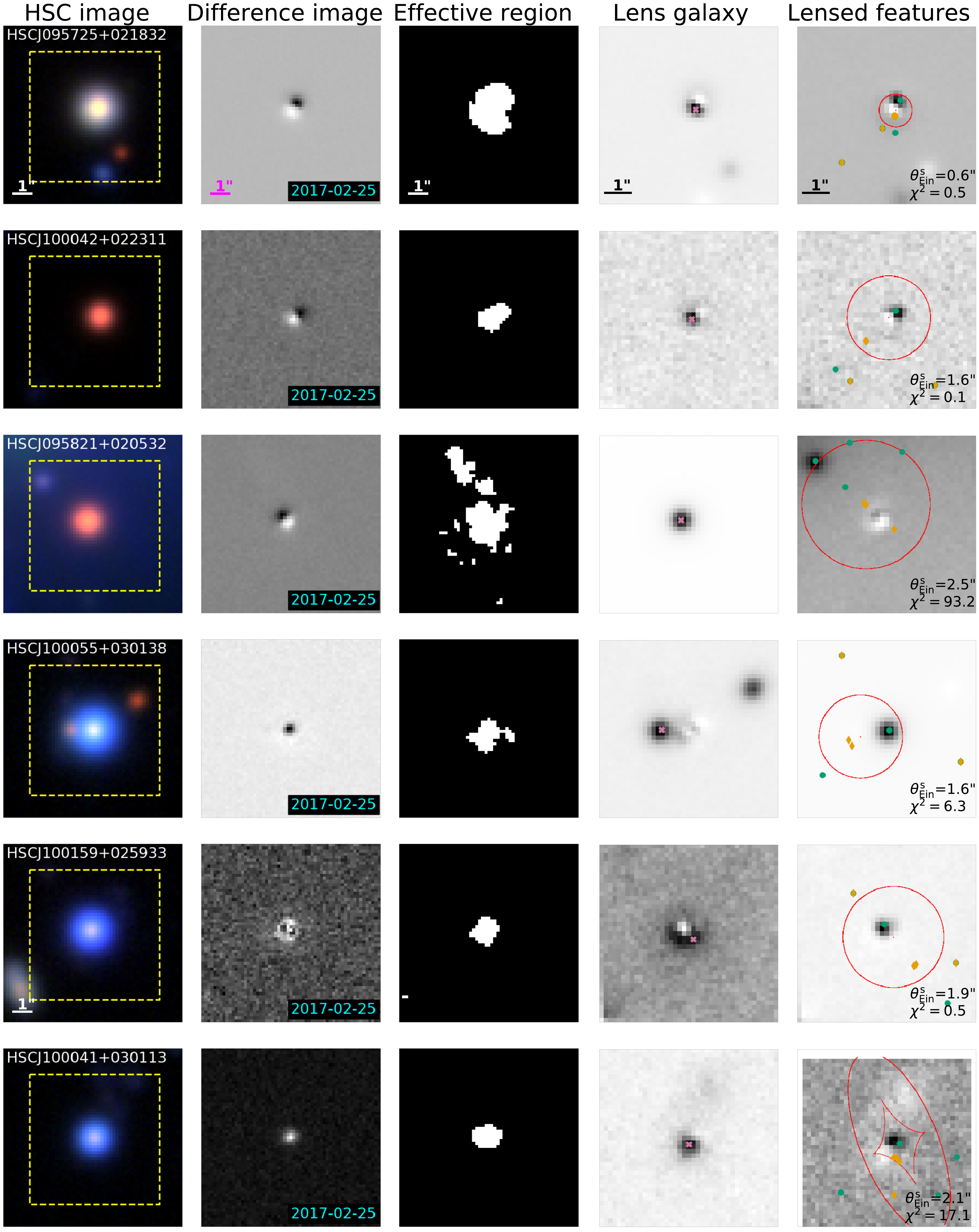}
\caption{Examples of variability false positives, the variable objects
  identified as lensed quasars by the variability-based method,
  but rejected by \textsc{Chitah}. The columns are in the same format as in Fig.~\ref{fig:final_candidates}.
  %
%
 See Sec.~\ref{sec:chitah_fp} for more details.} 
%
%
\label{fig:chitah_fp}
\end{figure*}

Many variability false positives that are point-like have effective region
areas that are even larger than most of our final candidates and the
\textsc{Chitah} false positives, which means that we take the risk of losing 
promising lensed quasar candidates if we discard these variability
false positives by applying stricter constraints on $p_\text{thrs}\%$ and $N_\text{thrs}$.
We show some examples of these point-like variability
false positives in Fig.~\ref{fig:chitah_fp}. 
In particular, HSCJ095725+021832, HSCJ100042+022311, and
HSCJ095821+020532 have `Taiji-like' dipole patterns -- half black and half white
in the difference images. The Taiji-like
patterns are possibly caused by the object brightnesses. 
From the simulation work done in C20, we found that an object starts to have the
Taiji-like dipole residual in the difference images when it is
brighter than $\lesssim$ 21.5 mag, regardless of whether the object is
variable. The possible reason is that the transient pipeline no
longer performs the image subtraction perfectly and it could not
subtract the light correctly when an object is too bright
($\lesssim$21.5 mag). The brighter the object is, the more obvious the
Taiji-like 
pattern is in the difference image. The Taiji-like residuals
contribute larger effective region areas, and the
variability-based selection would therefore misidentify these
point-like objects as lensed quasar candidates. We note that,
proper motions and parallaxes can also produce the Taiji-like pattern. While
this effect should be negligible in the HSC transient survey, it will start to
be visible in the 10-year survey of the LSST. Nonetheless, the Taiji-like pattern associated with proper
motions and parallaxes of stars would change over time, and the multiple epochs of difference images in the LSST
can help distinguish between proper motions of stars and the difference-image issue associated with bright objects.

In addition to those objects with Taiji-like residuals, some of the
variability false positives could also be variable stars or unlensed
quasars, such as HSCJ100159+025933 and HSCJ100041+030113.
They are
classified as a star and a quasar in the internal HSC transient catalogue,
respectively. Although not listed in the internal HSC transient catalogue, HSCJ100055+030138 is possibly a variable star or an unlensed quasar as well.
We inspected the difference images of 13 epochs, the corresponding effective regions, and the colour-composite images of HSCJ100055+030138 and found that its effective region areas might come from a substantial brightness change, instead of noise peaks or improper image subtraction. 
Ideally, lensed quasars have larger effective region
areas due to their multiple images, and strict thresholds on
the effective region areas for the variability selection should
largely decrease the number of those point-like false positives, such
as variable stars or unlensed quasars, without aggressively losing the
lensed quasar candidates. However, in this work, our final candidates
of lensed quasars generally have smaller effective regions than the
\textsc{Chitah} false positives and the variability
false positives. Although the loose constraints on the
variability-based selection in this work yield a huge number of
non-lensed objects, they seem to be necessary for securing the lensed
quasar candidates. Therefore, \textsc{Chitah} is important as the
second step in this lens search starting from time variability
since it discards the point-like objects with large effective regions
and efficiently decreases the number of lensed quasar candidates
without missing the promising ones.


\subsection{Discussion}
\label{sec:discussion}
We searched the literature to check if any of our final candidates have been found as a lens system.
Except for HSCJ095921+020638, none of the other final candidates have been found as a lens.
Among the final candidates, HSCJ100050+013251, HSCJ100307+020241, HSCJ095943+022046, and HSCJ095744+023835 
are listed in the galaxy catalogue of \citet{cosmos_gal_2007}, 
and our serendipitous discovery, HSCJ100129+024427, is listed as a bright galaxy in the catalogue of \citet{cosmos_gal_2009}. We also checked the WISE C75/R90 catalogues from \citet{c_79_r_90} and the Milliquas catalogue v7.1 (2021) \citep{milliquas}. For C75 and R90, none of our final candidates are included. For Milliquas, only the known quad, HSCJ095921+020638, is included.\\  

The recovery of the known lensed quasar, HSCJ095921+020638, and the
discovery of the other final candidates can represent the
completeness of our lensed quasar search, since we estimate that the number of quad(s) lying 
in the UltraDeep layer of the COSMOS field within the HSC transient
survey (\textit{i} $\sim$26.0 mag) is about $1\pm1$ based on the
lensing rates in OM10 and the correction in
\citet{Spinello_gaia_kids}. We further found that the strictest
criterion for HSCJ095921+020638 to qualify the variability-based
selection is $(p_\text{thrs}\%, N_\text{thrs})=(50\%, 6)$. With
this last criterion, the
variability-based selection would identify 32,127 out of the
101,353 HSC variables as lensed quasar candidates, and 936 out of these
32,127 objects would further be identified as lensed quasar candidates
by \textsc{Chitah}. In the end, under the same visual inspection, only
four objects in Table~\ref{tab:candidate} remain as final
candidates: HSCJ095921+020638, HSCJ095943+022046, HSCJ095744+023835,
and HSCJ100050+031825. We
note that the COSMOS field is small and HSCJ095921+020638 is faint, so the
loose constraints applied in this work to recover HSCJ095921+020638
might not be mandatory for a more general lensed quasar search in a
larger field. In fact, C20 has demonstrated that we could find bright lensed
quasars with a wide separation at true-positive rate of 90.1\% and
false-positive rate of 2.3\% with $(p_\text{thrs}\%,
N_\text{thrs})=(95\%, 9)$.  
In future applications of this method to larger cadenced image surveys,
the selection criteria are adjustable to balance between purity and
completeness.

\section{Conclusion}
\label{sec:conclusion}

In this work, we performed a lensed quasar search within the HSC
transient survey through the time variability of lensed quasars. We
first used \textit{i}-band difference images from the HSC transient
survey to select objects with the variability-based method in
C20 based on their spatial extent in the
difference image. We ran \textsc{Chitah}, a lens search algorithm
based on the image configuration, on the objects selected by the
difference image. Finally, we visually inspected the objects that were
both selected by the variability-based method and \textsc{Chitah}
and graded them. We summarise the work as follows:

\begin{enumerate}
\item Among the 101,353 HSC variables, 
  the variability-based method conservatively selected 83,657 HSC
  variables as potential lensed quasar candidates with loose
  criteria. Among the 83,657 HSC variables, \textsc{Chitah} further
  identified 2,130 as lensed quasar candidates. The visual inspection
  picked out seven from the 2,130 lensed quasar candidates as final
  candidates. In addition to the seven final candidates, we
  serendipitously found one lensed galaxy candidate.

\item As a first application, we used the variability-based method
  from C20 in a conservative manner. Although C20 method is helpful in selecting
  potential lensed quasar candidates by picking out objects with a
  larger spatial extent in the difference image, this work shows
  that our final candidates generally have a smaller spatial extent
  than the false positives, indicating that tightened
  criteria in the variability selection might not be an ideal means to
  improve the lens search efficiency in COSMOS.

\item Since using the variability-based method alone might not be
  efficient, \textsc{Chitah} is important as a further examination to
  remove the false positives from the variability selection through
  object configurations, especially for discarding those point-like
  false positives.

\item For future lensed quasar searches in larger sky areas, we could
  use stricter criteria for the variability-based method to
  reduce the false-positive rates. The only known lensed quasar in the
  field we have examined in this work, HSCJ095921+020638, is a special case
  of faint lensed quasars, and the loose constraints on the
  variability-based method exploited in this work to recover
  HSCJ095921+020638 are not indispensable.
\end{enumerate}

As this work has shown, the variability-based lens search method
from C20 is workable and could be applied to other cadenced imaging
surveys, and a combination with other lens search techniques as an
advanced check such as \textsc{Chitah} will improve the lens search
efficiency. The upcoming LSST is expected to have a difference imaging
process and image quality similar to the HSC transient survey, but
covering a much larger sky area. Therefore, we expect to discover new lensed quasars
in the LSST through variability-based searches.

\begin{acknowledgements}

  We thank Anupreeta More and Masaomi Tanaka for useful discussions. 
  We thank anonymous referee for helpful comments that improved the 
  presentation of our paper. DCYC thanks Yen-Ting Lin and Wei-Hao Wang
  for the hospitality at the Institute of Astronomy and Astrophysics, Academia Sinica (ASIAA).
  DCYC and SHS thank the Max Planck Society for support through the
  Max Planck Research Group for SHS.
  JHHC acknowledges support from the
  Swiss National Science Foundation (SNSF). ATJ is supported in part by JSPS KAKENHI Grant Number JP17H02868.
  This research made use of Astropy,\footnote{http://www.astropy.org}
  a community-developed core Python package for Astronomy
  \citep{astropy:2013, astropy:2018}. The Hyper Suprime-Cam (HSC) collaboration includes the astronomical communities of Japan and Taiwan, and Princeton University.  The HSC instrumentation and software were developed by the National Astronomical Observatory of Japan (NAOJ), the Kavli Institute for the Physics and Mathematics of the Universe (Kavli IPMU), the University of Tokyo, the High Energy Accelerator Research Organization (KEK), the Academia Sinica Institute for Astronomy and Astrophysics in Taiwan (ASIAA), and Princeton University.  Funding was contributed by the FIRST program from the Japanese Cabinet Office, the Ministry of Education, Culture, Sports, Science and Technology (MEXT), the Japan Society for the Promotion of Science (JSPS), Japan Science and Technology Agency  (JST), the Toray Science  Foundation, NAOJ, Kavli IPMU, KEK, ASIAA, and Princeton University. This paper makes use of software developed for the Large Synoptic Survey Telescope. We thank the LSST Project for making their code available as free software at http://dm.lsst.org. This paper is based on data collected at the Subaru Telescope and retrieved from the HSC data archive system, which is operated by Subaru Telescope and Astronomy Data Center (ADC) at NAOJ. Data analysis was in part carried out with the cooperation of Center for Computational Astrophysics (CfCA), NAOJ. The Pan-STARRS1 Surveys (PS1) and the PS1 public science archive have been made possible through contributions by the Institute for Astronomy, the University of Hawaii, the Pan-STARRS Project Office, the Max Planck Society and its participating institutes, the Max Planck Institute for Astronomy, Heidelberg, and the Max Planck Institute for Extraterrestrial Physics, Garching, The Johns Hopkins University, Durham University, the University of Edinburgh, the Queen’s University Belfast, the Harvard-Smithsonian Center for Astrophysics, the Las Cumbres Observatory Global Telescope Network Incorporated, the National Central University of Taiwan, the Space Telescope Science Institute, the National Aeronautics and Space Administration under grant No. NNX08AR22G issued through the Planetary Science Division of the NASA Science Mission Directorate, the National Science Foundation grant No. AST-1238877, the University of Maryland, Eotvos Lorand University (ELTE), the Los Alamos National Laboratory, and the Gordon and Betty Moore Foundation.

\end{acknowledgements}

\bibliographystyle{aa} 
\bibliography{HSC_var_lens.bib}

\begin{thebibliography}{73}
\expandafter\ifx\csname natexlab\endcsname\relax\def\natexlab#1{#1}\fi

\bibitem[{{Agnello}(2017)}]{Adriano_alone}
{Agnello}, A. 2017, \mnras, 471, 2013

\bibitem[{{Agnello} {et~al.}(2015){Agnello}, {Kelly}, {Treu}, \&
  {Marshall}}]{Adriano_method}
{Agnello}, A., {Kelly}, B.~C., {Treu}, T., \& {Marshall}, P.~J. 2015, \mnras,
  448, 1446

\bibitem[{{Agnello} {et~al.}(2018){Agnello}, {Lin}, {Kuropatkin},
  {Buckley-Geer}, {Anguita}, {Schechter}, {Morishita}, {Motta}, {Rojas},
  {Treu}, {Amara}, {Auger}, {Courbin}, {Fassnacht}, {Frieman}, {More},
  {Marshall}, {McMahon}, {Meylan}, {Suyu}, {Glazebrook}, {Morgan}, {Nord},
  {Abbott}, {Abdalla}, {Annis}, {Bechtol}, {Benoit-L{\'e}vy}, {Bertin},
  {Bernstein}, {Brooks}, {Burke}, {Rosell}, {Carretero}, {Cunha}, {D'Andrea},
  {da Costa}, {Desai}, {Drlica-Wagner}, {Eifler}, {Flaugher},
  {Garc{\'{\i}}a-Bellido}, {Gaztanaga}, {Gerdes}, {Gruen}, {Gruendl},
  {Gschwend}, {Gutierrez}, {Honscheid}, {James}, {Kuehn}, {Lahav}, {Lima},
  {Maia}, {March}, {Menanteau}, {Miquel}, {Ogando}, {Plazas}, {Sanchez},
  {Scarpine}, {Schindler}, {Schubnell}, {Sevilla-Noarbe}, {Smith},
  {Soares-Santos}, {Sobreira}, {Suchyta}, {Swanson}, {Tarle}, {Tucker}, \&
  {Wechsler}}]{Adriano_gaia_des_02}
{Agnello}, A., {Lin}, H., {Kuropatkin}, N., {et~al.} 2018, \mnras, 479, 4345

\bibitem[{{Agnello} \& {Spiniello}(2019)}]{Adriano_gaia_des_01}
{Agnello}, A. \& {Spiniello}, C. 2019, \mnras, 489, 2525

\bibitem[{{Aihara} {et~al.}(2018){Aihara}, {Arimoto}, {Armstrong}, {Arnouts},
  {Bahcall}, {Bickerton}, {Bosch}, {Bundy}, {Capak}, {Chan}, {Chiba}, {Coupon},
  {Egami}, {Enoki}, {Finet}, {Fujimori}, {Fujimoto}, {Furusawa}, {Furusawa},
  {Goto}, {Goulding}, {Greco}, {Greene}, {Gunn}, {Hamana}, {Harikane},
  {Hashimoto}, {Hattori}, {Hayashi}, {Hayashi}, {He{\l}miniak}, {Higuchi},
  {Hikage}, {Ho}, {Hsieh}, {Huang}, {Huang}, {Ikeda}, {Imanishi}, {Inoue},
  {Iwasawa}, {Iwata}, {Jaelani}, {Jian}, {Kamata}, {Karoji}, {Kashikawa},
  {Katayama}, {Kawanomoto}, {Kayo}, {Koda}, {Koike}, {Kojima}, {Komiyama},
  {Konno}, {Koshida}, {Koyama}, {Kusakabe}, {Leauthaud}, {Lee}, {Lin}, {Lin},
  {Lupton}, {Mandelbaum}, {Matsuoka}, {Medezinski}, {Mineo}, {Miyama},
  {Miyatake}, {Miyazaki}, {Momose}, {More}, {More}, {Moritani}, {Moriya},
  {Morokuma}, {Mukae}, {Murata}, {Murayama}, {Nagao}, {Nakata}, {Niida},
  {Niikura}, {Nishizawa}, {Obuchi}, {Oguri}, {Oishi}, {Okabe}, {Okamoto},
  {Okura}, {Ono}, {Onodera}, {Onoue}, {Osato}, {Ouchi}, {Price}, {Pyo}, {Sako},
  {Sawicki}, {Shibuya}, {Shimasaku}, {Shimono}, {Shirasaki}, {Silverman},
  {Simet}, {Speagle}, {Spergel}, {Strauss}, {Sugahara}, {Sugiyama}, {Suto},
  {Suyu}, {Suzuki}, {Tait}, {Takada}, {Takata}, {Tamura}, {Tanaka}, {Tanaka},
  {Tanaka}, {Tanaka}, {Terai}, {Terashima}, {Toba}, {Tominaga}, {Toshikawa},
  {Turner}, {Uchida}, {Uchiyama}, {Umetsu}, {Uraguchi}, {Urata}, {Usuda},
  {Utsumi}, {Wang}, {Wang}, {Wong}, {Yabe}, {Yamada}, {Yamanoi}, {Yasuda},
  {Yeh}, {Yonehara}, \& {Yuma}}]{Aihara_SSP}
{Aihara}, H., {Arimoto}, N., {Armstrong}, R., {et~al.} 2018, \pasj, 70, S4

\bibitem[{{Alard}(2000)}]{Alard_diff}
{Alard}, C. 2000, \aaps, 144, 363

\bibitem[{{Alard} \& {Lupton}(1998)}]{Robert_diff}
{Alard}, C. \& {Lupton}, R.~H. 1998, \apj, 503, 325

\bibitem[{{Anguita} {et~al.}(2009){Anguita}, {Faure}, {Kneib}, {Wambsganss},
  {Knobel}, {Koekemoer}, \& {Limousin}}]{Anguita_lens}
{Anguita}, T., {Faure}, C., {Kneib}, J.~P., {et~al.} 2009, \aap, 507, 35

\bibitem[{{Assef} {et~al.}(2018){Assef}, {Stern}, {Noirot}, {Jun}, {Cutri}, \&
  {Eisenhardt}}]{c_79_r_90}
{Assef}, R.~J., {Stern}, D., {Noirot}, G., {et~al.} 2018, \apjs, 234, 23

\bibitem[{{Astropy Collaboration} {et~al.}(2018){Astropy Collaboration},
  {Price-Whelan}, {Sip{\H o}cz}, {G{\"u}nther}, {Lim}, {Crawford}, {Conseil},
  {Shupe}, {Craig}, {Dencheva}, {Ginsburg}, {VanderPlas}, {Bradley},
  {P{\'e}rez-Su{\'a}rez}, {de Val-Borro}, {Aldcroft}, {Cruz}, {Robitaille},
  {Tollerud}, {Ardelean}, {Babej}, {Bach}, {Bachetti}, {Bakanov}, {Bamford},
  {Barentsen}, {Barmby}, {Baumbach}, {Berry}, {Biscani}, {Boquien}, {Bostroem},
  {Bouma}, {Brammer}, {Bray}, {Breytenbach}, {Buddelmeijer}, {Burke},
  {Calderone}, {Cano Rodr{\'{\i}}guez}, {Cara}, {Cardoso}, {Cheedella},
  {Copin}, {Corrales}, {Crichton}, {D'Avella}, {Deil}, {Depagne}, {Dietrich},
  {Donath}, {Droettboom}, {Earl}, {Erben}, {Fabbro}, {Ferreira}, {Finethy},
  {Fox}, {Garrison}, {Gibbons}, {Goldstein}, {Gommers}, {Greco}, {Greenfield},
  {Groener}, {Grollier}, {Hagen}, {Hirst}, {Homeier}, {Horton}, {Hosseinzadeh},
  {Hu}, {Hunkeler}, {Ivezi{\'c}}, {Jain}, {Jenness}, {Kanarek}, {Kendrew},
  {Kern}, {Kerzendorf}, {Khvalko}, {King}, {Kirkby}, {Kulkarni}, {Kumar},
  {Lee}, {Lenz}, {Littlefair}, {Ma}, {Macleod}, {Mastropietro}, {McCully},
  {Montagnac}, {Morris}, {Mueller}, {Mumford}, {Muna}, {Murphy}, {Nelson},
  {Nguyen}, {Ninan}, {N{\"o}the}, {Ogaz}, {Oh}, {Parejko}, {Parley}, {Pascual},
  {Patil}, {Patil}, {Plunkett}, {Prochaska}, {Rastogi}, {Reddy Janga},
  {Sabater}, {Sakurikar}, {Seifert}, {Sherbert}, {Sherwood-Taylor}, {Shih},
  {Sick}, {Silbiger}, {Singanamalla}, {Singer}, {Sladen}, {Sooley},
  {Sornarajah}, {Streicher}, {Teuben}, {Thomas}, {Tremblay}, {Turner},
  {Terr{\'o}n}, {van Kerkwijk}, {de la Vega}, {Watkins}, {Weaver}, {Whitmore},
  {Woillez}, {Zabalza}, \& {Astropy Contributors}}]{astropy:2018}
{Astropy Collaboration}, {Price-Whelan}, A.~M., {Sip{\H o}cz}, B.~M., {et~al.}
  2018, \aj, 156, 123

\bibitem[{{Astropy Collaboration} {et~al.}(2013){Astropy Collaboration},
  {Robitaille}, {Tollerud}, {Greenfield}, {Droettboom}, {Bray}, {Aldcroft},
  {Davis}, {Ginsburg}, {Price-Whelan}, {Kerzendorf}, {Conley}, {Crighton},
  {Barbary}, {Muna}, {Ferguson}, {Grollier}, {Parikh}, {Nair}, {Unther},
  {Deil}, {Woillez}, {Conseil}, {Kramer}, {Turner}, {Singer}, {Fox}, {Weaver},
  {Zabalza}, {Edwards}, {Azalee Bostroem}, {Burke}, {Casey}, {Crawford},
  {Dencheva}, {Ely}, {Jenness}, {Labrie}, {Lim}, {Pierfederici}, {Pontzen},
  {Ptak}, {Refsdal}, {Servillat}, \& {Streicher}}]{astropy:2013}
{Astropy Collaboration}, {Robitaille}, T.~P., {Tollerud}, E.~J., {et~al.} 2013,
  \aap, 558, A33

\bibitem[{{Browne} {et~al.}(2003){Browne}, {Wilkinson}, {Jackson}, {Myers},
  {Fassnacht}, {Koopmans}, {Marlow}, {Norbury}, {Rusin}, {Sykes}, {Biggs},
  {Blandford}, {de Bruyn}, {Chae}, {Helbig}, {King}, {McKean}, {Pearson},
  {Phillips}, {Readhead}, {Xanthopoulos}, \& {York}}]{class_02}
{Browne}, I.~W.~A., {Wilkinson}, P.~N., {Jackson}, N.~J.~F., {et~al.} 2003,
  \mnras, 341, 13

\bibitem[{{Capak} {et~al.}(2007){Capak}, {Aussel}, {Ajiki}, {McCracken},
  {Mobasher}, {Scoville}, {Shopbell}, {Taniguchi}, {Thompson}, {Tribiano},
  {Sasaki}, {Blain}, {Brusa}, {Carilli}, {Comastri}, {Carollo}, {Cassata},
  {Colbert}, {Ellis}, {Elvis}, {Giavalisco}, {Green}, {Guzzo}, {Hasinger},
  {Ilbert}, {Impey}, {Jahnke}, {Kartaltepe}, {Kneib}, {Koda}, {Koekemoer},
  {Komiyama}, {Leauthaud}, {Le Fevre}, {Lilly}, {Liu}, {Massey}, {Miyazaki},
  {Murayama}, {Nagao}, {Peacock}, {Pickles}, {Porciani}, {Renzini}, {Rhodes},
  {Rich}, {Salvato}, {Sanders}, {Scarlata}, {Schiminovich}, {Schinnerer},
  {Scodeggio}, {Sheth}, {Shioya}, {Tasca}, {Taylor}, {Yan}, \&
  {Zamorani}}]{cosmos_gal_2007}
{Capak}, P., {Aussel}, H., {Ajiki}, M., {et~al.} 2007, \apjs, 172, 99

\bibitem[{{Capelo} {et~al.}(2015){Capelo}, {Volonteri}, {Dotti}, {Bellovary},
  {Mayer}, \& {Governato}}]{capelo_merger}
{Capelo}, P.~R., {Volonteri}, M., {Dotti}, M., {et~al.} 2015, \mnras, 447, 2123

\bibitem[{{Chambers} {et~al.}(2016){Chambers}, {Magnier}, {Metcalfe},
  {Flewelling}, {Huber}, {Waters}, {Denneau}, {Draper}, {Farrow}, {Finkbeiner},
  {Holmberg}, {Koppenhoefer}, {Price}, {Rest}, {Saglia}, {Schlafly}, {Smartt},
  {Sweeney}, {Wainscoat}, {Burgett}, {Chastel}, {Grav}, {Heasley}, {Hodapp},
  {Jedicke}, {Kaiser}, {Kudritzki}, {Luppino}, {Lupton}, {Monet}, {Morgan},
  {Onaka}, {Shiao}, {Stubbs}, {Tonry}, {White}, {Ba{\~n}ados}, {Bell},
  {Bender}, {Bernard}, {Boegner}, {Boffi}, {Botticella}, {Calamida},
  {Casertano}, {Chen}, {Chen}, {Cole}, {Deacon}, {Frenk}, {Fitzsimmons},
  {Gezari}, {Gibbs}, {Goessl}, {Goggia}, {Gourgue}, {Goldman}, {Grant},
  {Grebel}, {Hambly}, {Hasinger}, {Heavens}, {Heckman}, {Henderson}, {Henning},
  {Holman}, {Hopp}, {Ip}, {Isani}, {Jackson}, {Keyes}, {Koekemoer}, {Kotak},
  {Le}, {Liska}, {Long}, {Lucey}, {Liu}, {Martin}, {Masci}, {McLean}, {Mindel},
  {Misra}, {Morganson}, {Murphy}, {Obaika}, {Narayan}, {Nieto-Santisteban},
  {Norberg}, {Peacock}, {Pier}, {Postman}, {Primak}, {Rae}, {Rai}, {Riess},
  {Riffeser}, {Rix}, {R{\"o}ser}, {Russel}, {Rutz}, {Schilbach}, {Schultz},
  {Scolnic}, {Strolger}, {Szalay}, {Seitz}, {Small}, {Smith}, {Soderblom},
  {Taylor}, {Thomson}, {Taylor}, {Thakar}, {Thiel}, {Thilker}, {Unger},
  {Urata}, {Valenti}, {Wagner}, {Walder}, {Walter}, {Watters}, {Werner},
  {Wood-Vasey}, \& {Wyse}}]{panstarrs_survey}
{Chambers}, K.~C., {Magnier}, E.~A., {Metcalfe}, N., {et~al.} 2016, arXiv
  e-prints [\eprint[arXiv]{1612.05560}]

\bibitem[{{Chan} {et~al.}(2015){Chan}, {Suyu}, {Chiueh}, {More}, {Marshall},
  {Coupon}, {Oguri}, \& {Price}}]{chitah}
{Chan}, J.~H.~H., {Suyu}, S.~H., {Chiueh}, T., {et~al.} 2015, \apj, 807, 138

\bibitem[{{Chan} {et~al.}(2020){Chan}, {Suyu}, {Sonnenfeld}, {Jaelani}, {More},
  {Yonehara}, {Kubota}, {Coupon}, {Lee}, {Oguri}, {Rusu}, \&
  {Wong}}]{chitah_hsc}
{Chan}, J. H.~H., {Suyu}, S.~H., {Sonnenfeld}, A., {et~al.} 2020, \aap, 636,
  A87

\bibitem[{{Chao} {et~al.}(2020){Chao}, {Chan}, {Suyu}, {Yasuda}, {More},
  {Oguri}, {Morokuma}, \& {Jaelani}}]{DC_Buddy}
{Chao}, D. C.~Y., {Chan}, J. H.~H., {Suyu}, S.~H., {et~al.} 2020, \aap, 640,
  A88

\bibitem[{{Chen} {et~al.}(2019){Chen}, {Fassnacht}, {Suyu}, {Rusu}, {Chan},
  {Wong}, {Auger}, {Hilbert}, {Bonvin}, {Birrer}, {Millon}, {Koopmans},
  {Lagattuta}, {McKean}, {Vegetti}, {Courbin}, {Ding}, {Halkola}, {Jee},
  {Shajib}, {Sluse}, {Sonnenfeld}, \& {Treu}}]{holicow_h0_02}
{Chen}, G. C.~F., {Fassnacht}, C.~D., {Suyu}, S.~H., {et~al.} 2019, \mnras,
  490, 1743

\bibitem[{{Comerford} {et~al.}(2015){Comerford}, {Pooley}, {Barrows}, {Greene},
  {Zakamska}, {Madejski}, \& {Cooper}}]{commerford_merger}
{Comerford}, J.~M., {Pooley}, D., {Barrows}, R.~S., {et~al.} 2015, \apj, 806,
  219

\bibitem[{{Delchambre} {et~al.}(2019){Delchambre}, {Krone-Martins}, {Wertz},
  {Ducourant}, {Galluccio}, {Kl{\"u}ter}, {Mignard}, {Teixeira}, {Djorgovski},
  {Stern}, {Graham}, {Surdej}, {Bastian}, {Wambsganss}, {Le Campion}, \&
  {Slezak}}]{gaia_ml}
{Delchambre}, L., {Krone-Martins}, A., {Wertz}, O., {et~al.} 2019, \aap, 622,
  A165

\bibitem[{{Ding} {et~al.}(2017){Ding}, {Treu}, {Suyu}, {Wong}, {Morishita},
  {Park}, {Sluse}, {Auger}, {Agnello}, {Bennert}, \&
  {Collett}}]{quasar_host_02}
{Ding}, X., {Treu}, T., {Suyu}, S.~H., {et~al.} 2017, \mnras, 472, 90

\bibitem[{{Drake} {et~al.}(2009){Drake}, {Djorgovski}, {Mahabal}, {Beshore},
  {Larson}, {Graham}, {Williams}, {Christensen}, {Catelan}, {Boattini},
  {Gibbs}, {Hill}, \& {Kowalski}}]{CRTS}
{Drake}, A.~J., {Djorgovski}, S.~G., {Mahabal}, A., {et~al.} 2009, \apj, 696,
  870

\bibitem[{{Ellison} {et~al.}(2011){Ellison}, {Patton}, {Mendel}, \&
  {Scudder}}]{ellison_merger}
{Ellison}, S.~L., {Patton}, D.~R., {Mendel}, J.~T., \& {Scudder}, J.~M. 2011,
  \mnras, 418, 2043

\bibitem[{{Fan} {et~al.}(2019){Fan}, {Wang}, {Yang}, {Keeton}, {Yue},
  {Zabludoff}, {Bian}, {Bonaglia}, {Georgiev}, {Hennawi}, {Li}, {McGreer},
  {Naidu}, {Pacucci}, {Rabien}, {Thompson}, {Venemans}, {Walter}, {Wang}, \&
  {Wu}}]{high_z_lensed_quasar}
{Fan}, X., {Wang}, F., {Yang}, J., {et~al.} 2019, \apjl, 870, L11

\bibitem[{{Flesch}(2019)}]{milliquas}
{Flesch}, E.~W. 2019, arXiv e-prints, arXiv:1912.05614

\bibitem[{{Furusawa} {et~al.}(2018){Furusawa}, {Koike}, {Takata}, {Okura},
  {Miyatake}, {Lupton}, {Bickerton}, {Price}, {Bosch}, {Yasuda}, {Mineo},
  {Yamada}, {Miyazaki}, {Nakata}, {Koshida}, {Komiyama}, {Utsumi},
  {Kawanomoto}, {Jeschke}, {Noumaru}, {Schubert}, {Iwata}, {Finet},
  {Fujiyoshi}, {Tajitsu}, {Terai}, \& {Lee}}]{HSC_Furusawa_2018}
{Furusawa}, H., {Koike}, M., {Takata}, T., {et~al.} 2018, \pasj, 70, S3

\bibitem[{{Gaia Collaboration} {et~al.}(2016){Gaia Collaboration}, {Prusti},
  {de Bruijne}, {Brown}, {Vallenari}, {Babusiaux}, {Bailer-Jones}, {Bastian},
  {Biermann}, {Evans}, \& et~al.}]{Gaia_survey}
{Gaia Collaboration}, {Prusti}, T., {de Bruijne}, J.~H.~J., {et~al.} 2016,
  \aap, 595, A1

\bibitem[{{Gilman} {et~al.}(2019){Gilman}, {Birrer}, {Treu}, {Nierenberg}, \&
  {Benson}}]{Gilman_Sub}
{Gilman}, D., {Birrer}, S., {Treu}, T., {Nierenberg}, A., \& {Benson}, A. 2019,
  \mnras, 487, 5721

\bibitem[{{Ilbert} {et~al.}(2009){Ilbert}, {Capak}, {Salvato}, {Aussel},
  {McCracken}, {Sanders}, {Scoville}, {Kartaltepe}, {Arnouts}, {Le Floc'h},
  {Mobasher}, {Taniguchi}, {Lamareille}, {Leauthaud}, {Sasaki}, {Thompson},
  {Zamojski}, {Zamorani}, {Bardelli}, {Bolzonella}, {Bongiorno}, {Brusa},
  {Caputi}, {Carollo}, {Contini}, {Cook}, {Coppa}, {Cucciati}, {de la Torre},
  {de Ravel}, {Franzetti}, {Garilli}, {Hasinger}, {Iovino}, {Kampczyk},
  {Kneib}, {Knobel}, {Kovac}, {Le Borgne}, {Le Brun}, {Le F{\`e}vre}, {Lilly},
  {Looper}, {Maier}, {Mainieri}, {Mellier}, {Mignoli}, {Murayama}, {Pell{\`o}},
  {Peng}, {P{\'e}rez-Montero}, {Renzini}, {Ricciardelli}, {Schiminovich},
  {Scodeggio}, {Shioya}, {Silverman}, {Surace}, {Tanaka}, {Tasca}, {Tresse},
  {Vergani}, \& {Zucca}}]{cosmos_source}
{Ilbert}, O., {Capak}, P., {Salvato}, M., {et~al.} 2009, \apj, 690, 1236

\bibitem[{{Inada} {et~al.}(2008){Inada}, {Oguri}, {Becker}, {Shin}, {Richards},
  {Hennawi}, {White}, {Pindor}, {Strauss}, {Kochanek}, {Johnston}, {Gregg},
  {Kayo}, {Eisenstein}, {Hall}, {Castander}, {Clocchiatti}, {Anderson},
  {Schneider}, {York}, {Lupton}, {Chiu}, {Kawano}, {Scranton}, {Frieman},
  {Keeton}, {Morokuma}, {Rix}, {Turner}, {Burles}, {Brunner}, {Sheldon},
  {Bahcall}, \& {Masataka}}]{sqls_02}
{Inada}, N., {Oguri}, M., {Becker}, R.~H., {et~al.} 2008, \aj, 135, 496

\bibitem[{{Inada} {et~al.}(2010){Inada}, {Oguri}, {Shin}, {Kayo}, {Strauss},
  {Hennawi}, {Morokuma}, {Becker}, {White}, {Kochanek}, {Gregg}, {Chiu},
  {Johnston}, {Clocchiatti}, {Richards}, {Schneider}, {Frieman}, {Fukugita},
  {Gott}, {Hall}, {York}, {Castander}, \& {Bahcall}}]{sqls_03}
{Inada}, N., {Oguri}, M., {Shin}, M.-S., {et~al.} 2010, \aj, 140, 403

\bibitem[{{Inada} {et~al.}(2012){Inada}, {Oguri}, {Shin}, {Kayo}, {Strauss},
  {Morokuma}, {Rusu}, {Fukugita}, {Kochanek}, {Richards}, {Schneider}, {York},
  {Bahcall}, {Frieman}, {Hall}, \& {White}}]{sqls_04}
{Inada}, N., {Oguri}, M., {Shin}, M.-S., {et~al.} 2012, \aj, 143, 119

\bibitem[{{Ivezi{\'c}} {et~al.}(2019){Ivezi{\'c}}, {Kahn}, {Tyson}, {Abel},
  {Acosta}, {Allsman}, {Alonso}, {AlSayyad}, {Anderson}, {Andrew}, {Angel},
  {Angeli}, {Ansari}, {Antilogus}, {Araujo}, {Armstrong}, {Arndt}, {Astier},
  {Aubourg}, {Auza}, {Axelrod}, {Bard}, {Barr}, {Barrau}, {Bartlett}, {Bauer},
  {Bauman}, {Baumont}, {Bechtol}, {Bechtol}, {Becker}, {Becla}, {Beldica},
  {Bellavia}, {Bianco}, {Biswas}, {Blanc}, {Blazek}, {Bland ford}, {Bloom},
  {Bogart}, {Bond}, {Booth}, {Borgland}, {Borne}, {Bosch}, {Boutigny},
  {Brackett}, {Bradshaw}, {Brand t}, {Brown}, {Bullock}, {Burchat}, {Burke},
  {Cagnoli}, {Calabrese}, {Callahan}, {Callen}, {Carlin}, {Carlson}, {Chand
  rasekharan}, {Charles-Emerson}, {Chesley}, {Cheu}, {Chiang}, {Chiang},
  {Chirino}, {Chow}, {Ciardi}, {Claver}, {Cohen-Tanugi}, {Cockrum}, {Coles},
  {Connolly}, {Cook}, {Cooray}, {Covey}, {Cribbs}, {Cui}, {Cutri}, {Daly},
  {Daniel}, {Daruich}, {Daubard}, {Daues}, {Dawson}, {Delgado}, {Dellapenna},
  {de Peyster}, {de Val-Borro}, {Digel}, {Doherty}, {Dubois},
  {Dubois-Felsmann}, {Durech}, {Economou}, {Eifler}, {Eracleous}, {Emmons},
  {Fausti Neto}, {Ferguson}, {Figueroa}, {Fisher-Levine}, {Focke}, {Foss},
  {Frank}, {Freemon}, {Gangler}, {Gawiser}, {Geary}, {Gee}, {Geha}, {Gessner},
  {Gibson}, {Gilmore}, {Glanzman}, {Glick}, {Goldina}, {Goldstein}, {Goodenow},
  {Graham}, {Gressler}, {Gris}, {Guy}, {Guyonnet}, {Haller}, {Harris},
  {Hascall}, {Haupt}, {Hernand ez}, {Herrmann}, {Hileman}, {Hoblitt},
  {Hodgson}, {Hogan}, {Howard}, {Huang}, {Huffer}, {Ingraham}, {Innes},
  {Jacoby}, {Jain}, {Jammes}, {Jee}, {Jenness}, {Jernigan}, {Jevremovi{\'c}},
  {Johns}, {Johnson}, {Johnson}, {Jones}, {Juramy-Gilles}, {Juri{\'c}},
  {Kalirai}, {Kallivayalil}, {Kalmbach}, {Kantor}, {Karst}, {Kasliwal},
  {Kelly}, {Kessler}, {Kinnison}, {Kirkby}, {Knox}, {Kotov}, {Krabbendam},
  {Krughoff}, {Kub{\'a}nek}, {Kuczewski}, {Kulkarni}, {Ku}, {Kurita}, {Lage},
  {Lambert}, {Lange}, {Langton}, {Le Guillou}, {Levine}, {Liang}, {Lim},
  {Lintott}, {Long}, {Lopez}, {Lotz}, {Lupton}, {Lust}, {MacArthur}, {Mahabal},
  {Mand elbaum}, {Markiewicz}, {Marsh}, {Marshall}, {Marshall}, {May},
  {McKercher}, {McQueen}, {Meyers}, {Migliore}, {Miller}, {Mills}, {Miraval},
  {Moeyens}, {Moolekamp}, {Monet}, {Moniez}, {Monkewitz}, {Montgomery},
  {Morrison}, {Mueller}, {Muller}, {Mu{\~n}oz Arancibia}, {Neill}, {Newbry},
  {Nief}, {Nomerotski}, {Nordby}, {O'Connor}, {Oliver}, {Olivier}, {Olsen},
  {O'Mullane}, {Ortiz}, {Osier}, {Owen}, {Pain}, {Palecek}, {Parejko},
  {Parsons}, {Pease}, {Peterson}, {Peterson}, {Petravick}, {Libby Petrick},
  {Petry}, {Pierfederici}, {Pietrowicz}, {Pike}, {Pinto}, {Plante}, {Plate},
  {Plutchak}, {Price}, {Prouza}, {Radeka}, {Rajagopal}, {Rasmussen},
  {Regnault}, {Reil}, {Reiss}, {Reuter}, {Ridgway}, {Riot}, {Ritz}, {Robinson},
  {Roby}, {Roodman}, {Rosing}, {Roucelle}, {Rumore}, {Russo}, {Saha},
  {Sassolas}, {Schalk}, {Schellart}, {Schindler}, {Schmidt}, {Schneider},
  {Schneider}, {Schoening}, {Schumacher}, {Schwamb}, {Sebag}, {Selvy},
  {Sembroski}, {Seppala}, {Serio}, {Serrano}, {Shaw}, {Shipsey}, {Sick},
  {Silvestri}, {Slater}, {Smith}, {Smith}, {Sobhani}, {Soldahl},
  {Storrie-Lombardi}, {Stover}, {Strauss}, {Street}, {Stubbs}, {Sullivan},
  {Sweeney}, {Swinbank}, {Szalay}, {Takacs}, {Tether}, {Thaler}, {Thayer},
  {Thomas}, {Thornton}, {Thukral}, {Tice}, {Trilling}, {Turri}, {Van Berg},
  {Vanden Berk}, {Vetter}, {Virieux}, {Vucina}, {Wahl}, {Walkowicz}, {Walsh},
  {Walter}, {Wang}, {Wang}, {Warner}, {Wiecha}, {Willman}, {Winters},
  {Wittman}, {Wolff}, {Wood-Vasey}, {Wu}, {Xin}, {Yoachim}, \&
  {Zhan}}]{lsst_survey}
{Ivezi{\'c}}, {\v{Z}}., {Kahn}, S.~M., {Tyson}, J.~A., {et~al.} 2019, \apj,
  873, 111

\bibitem[{{Jaelani} {et~al.}(2020){Jaelani}, {More}, {Oguri}, {Sonnenfeld},
  {Suyu}, {Rusu}, {Wong}, {Chan}, {Kayo}, {Lee}, {Chao}, {Coupon}, {Inoue}, \&
  {Futamase}}]{anton_hsc}
{Jaelani}, A.~T., {More}, A., {Oguri}, M., {et~al.} 2020, \mnras, 495, 1291

\bibitem[{{Kawanomoto} {et~al.}(2018){Kawanomoto}, {Uraguchi}, {Komiyama},
  {Miyazaki}, {Furusawa}, {Finet}, {Hattori}, {Wang}, {Yasuda}, \&
  {Suzuki}}]{HSC_Kawanomoto_2018}
{Kawanomoto}, S., {Uraguchi}, F., {Komiyama}, Y., {et~al.} 2018, \pasj, 70, 66

\bibitem[{{Kochanek} {et~al.}(2006){Kochanek}, {Mochejska}, {Morgan}, \&
  {Stanek}}]{Kochanek_method}
{Kochanek}, C.~S., {Mochejska}, B., {Morgan}, N.~D., \& {Stanek}, K.~Z. 2006,
  \apjl, 637, L73

\bibitem[{{Koekemoer} {et~al.}(2007){Koekemoer}, {Aussel}, {Calzetti}, {Capak},
  {Giavalisco}, {Kneib}, {Leauthaud}, {Le F{\`e}vre}, {McCracken}, {Massey},
  {Mobasher}, {Rhodes}, {Scoville}, \& {Shopbell}}]{hst_treasure}
{Koekemoer}, A.~M., {Aussel}, H., {Calzetti}, D., {et~al.} 2007, \apjs, 172,
  196

\bibitem[{{Komiyama} {et~al.}(2018){Komiyama}, {Obuchi}, {Nakaya}, {Kamata},
  {Kawanomoto}, {Utsumi}, {Miyazaki}, {Uraguchi}, {Furusawa}, {Morokuma},
  {Uchida}, {Miyatake}, {Mineo}, {Fujimori}, {Aihara}, {Karoji}, {Gunn}, \&
  {Wang}}]{HSC_Komiyama_2018}
{Komiyama}, Y., {Obuchi}, Y., {Nakaya}, H., {et~al.} 2018, \pasj, 70, S2

\bibitem[{{Koopmans} {et~al.}(2006){Koopmans}, {Treu}, {Bolton}, {Burles}, \&
  {Moustakas}}]{Koopman_centre}
{Koopmans}, L. V.~E., {Treu}, T., {Bolton}, A.~S., {Burles}, S., \&
  {Moustakas}, L.~A. 2006, \apj, 649, 599

\bibitem[{{Kormann} {et~al.}(1994){Kormann}, {Schneider}, \&
  {Bartelmann}}]{SIE}
{Kormann}, R., {Schneider}, P., \& {Bartelmann}, M. 1994, \aap, 284, 285

\bibitem[{{Krone-Martins} {et~al.}(2019){Krone-Martins}, {Graham}, {Stern},
  {Djorgovski}, {Delchambre}, {Ducourant}, {Teixeira}, {Drake}, {Scarano},
  {Surdej}, {Galluccio}, {Jalan}, {Wertz}, {Kl{\"u}ter}, {Mignard},
  {Spindola-Duarte}, {Dobie}, {Slezak}, {Sluse}, {Murphy}, {Boehm},
  {Nierenberg}, {Bastian}, {Wambsganss}, \& {LeCampion}}]{wise_gaia}
{Krone-Martins}, A., {Graham}, M.~J., {Stern}, D., {et~al.} 2019, arXiv
  e-prints, arXiv:1912.08977

\bibitem[{{Lemon} {et~al.}(2019){Lemon}, {Auger}, \& {McMahon}}]{Cameron_03}
{Lemon}, C.~A., {Auger}, M.~W., \& {McMahon}, R.~G. 2019, \mnras, 483, 4242

\bibitem[{{Lemon} {et~al.}(2017){Lemon}, {Auger}, {McMahon}, \&
  {Koposov}}]{Cameron_01}
{Lemon}, C.~A., {Auger}, M.~W., {McMahon}, R.~G., \& {Koposov}, S.~E. 2017,
  \mnras, 472, 5023

\bibitem[{{Lemon} {et~al.}(2018){Lemon}, {Auger}, {McMahon}, \&
  {Ostrovski}}]{Cameron_02}
{Lemon}, C.~A., {Auger}, M.~W., {McMahon}, R.~G., \& {Ostrovski}, F. 2018,
  \mnras, 479, 5060

\bibitem[{{Lilly} {et~al.}(2009){Lilly}, {Le Brun}, {Maier}, {Mainieri},
  {Mignoli}, {Scodeggio}, {Zamorani}, {Carollo}, {Contini}, {Kneib}, {Le
  F{\`e}vre}, {Renzini}, {Bardelli}, {Bolzonella}, {Bongiorno}, {Caputi},
  {Coppa}, {Cucciati}, {de la Torre}, {de Ravel}, {Franzetti}, {Garilli},
  {Iovino}, {Kampczyk}, {Kovac}, {Knobel}, {Lamareille}, {Le Borgne}, {Pello},
  {Peng}, {P{\'e}rez-Montero}, {Ricciardelli}, {Silverman}, {Tanaka}, {Tasca},
  {Tresse}, {Vergani}, {Zucca}, {Ilbert}, {Salvato}, {Oesch}, {Abbas},
  {Bottini}, {Capak}, {Cappi}, {Cassata}, {Cimatti}, {Elvis}, {Fumana},
  {Guzzo}, {Hasinger}, {Koekemoer}, {Leauthaud}, {Maccagni}, {Marinoni},
  {McCracken}, {Memeo}, {Meneux}, {Porciani}, {Pozzetti}, {Sanders},
  {Scaramella}, {Scarlata}, {Scoville}, {Shopbell}, \&
  {Taniguchi}}]{cosmos_gal_2009}
{Lilly}, S.~J., {Le Brun}, V., {Maier}, C., {et~al.} 2009, \apjs, 184, 218

\bibitem[{{Marshall} {et~al.}(2016){Marshall}, {Verma}, {More}, {Davis},
  {More}, {Kapadia}, {Parrish}, {Snyder}, {Wilcox}, {Baeten}, {Macmillan},
  {Cornen}, {Baumer}, {Simpson}, {Lintott}, {Miller}, {Paget}, {Simpson},
  {Smith}, {K{\"u}ng}, {Saha}, \& {Collett}}]{space_warp_01}
{Marshall}, P.~J., {Verma}, A., {More}, A., {et~al.} 2016, \mnras, 455, 1171

\bibitem[{{Miyazaki} {et~al.}(2018{\natexlab{a}}){Miyazaki}, {Komiyama},
  {Kawanomoto}, {Doi}, {Furusawa}, {Hamana}, {Hayashi}, {Ikeda}, {Kamata},
  {Karoji}, {Koike}, {Kurakami}, {Miyama}, {Morokuma}, {Nakata}, {Namikawa},
  {Nakaya}, {Nariai}, {Obuchi}, {Oishi}, {Okada}, {Okura}, {Tait}, {Takata},
  {Tanaka}, {Tanaka}, {Terai}, {Tomono}, {Uraguchi}, {Usuda}, {Utsumi},
  {Yamada}, {Yamanoi}, {Aihara}, {Fujimori}, {Mineo}, {Miyatake}, {Oguri},
  {Uchida}, {Tanaka}, {Yasuda}, {Takada}, {Murayama}, {Nishizawa}, {Sugiyama},
  {Chiba}, {Futamase}, {Wang}, {Chen}, {Ho}, {Liaw}, {Chiu}, {Ho}, {Lai},
  {Lee}, {Jeng}, {Iwamura}, {Armstrong}, {Bickerton}, {Bosch}, {Gunn},
  {Lupton}, {Loomis}, {Price}, {Smith}, {Strauss}, {Turner}, {Suzuki},
  {Miyazaki}, {Muramatsu}, {Yamamoto}, {Endo}, {Ezaki}, {Ito}, {Kawaguchi},
  {Sofuku}, {Taniike}, {Akutsu}, {Dojo}, {Kasumi}, {Matsuda}, {Imoto}, {Miwa},
  {Suzuki}, {Takeshi}, \& {Yokota}}]{Miyazaki_subaru}
{Miyazaki}, S., {Komiyama}, Y., {Kawanomoto}, S., {et~al.} 2018{\natexlab{a}},
  \pasj, 70, S1

\bibitem[{{Miyazaki} {et~al.}(2018{\natexlab{b}}){Miyazaki}, {Komiyama},
  {Kawanomoto}, {Doi}, {Furusawa}, {Hamana}, {Hayashi}, {Ikeda}, {Kamata},
  {Karoji}, {Koike}, {Kurakami}, {Miyama}, {Morokuma}, {Nakata}, {Namikawa},
  {Nakaya}, {Nariai}, {Obuchi}, {Oishi}, {Okada}, {Okura}, {Tait}, {Takata},
  {Tanaka}, {Tanaka}, {Terai}, {Tomono}, {Uraguchi}, {Usuda}, {Utsumi},
  {Yamada}, {Yamanoi}, {Aihara}, {Fujimori}, {Mineo}, {Miyatake}, {Oguri},
  {Uchida}, {Tanaka}, {Yasuda}, {Takada}, {Murayama}, {Nishizawa}, {Sugiyama},
  {Chiba}, {Futamase}, {Wang}, {Chen}, {Ho}, {Liaw}, {Chiu}, {Ho}, {Lai},
  {Lee}, {Jeng}, {Iwamura}, {Armstrong}, {Bickerton}, {Bosch}, {Gunn},
  {Lupton}, {Loomis}, {Price}, {Smith}, {Strauss}, {Turner}, {Suzuki},
  {Miyazaki}, {Muramatsu}, {Yamamoto}, {Endo}, {Ezaki}, {Ito}, {Kawaguchi},
  {Sofuku}, {Taniike}, {Akutsu}, {Dojo}, {Kasumi}, {Matsuda}, {Imoto}, {Miwa},
  {Suzuki}, {Takeshi}, \& {Yokota}}]{HSC_Miyazaki_2018}
{Miyazaki}, S., {Komiyama}, Y., {Kawanomoto}, S., {et~al.} 2018{\natexlab{b}},
  \pasj, 70, S1

\bibitem[{{Miyazaki} {et~al.}(2012){Miyazaki}, {Komiyama}, {Nakaya}, {Kamata},
  {Doi}, {Hamana}, {Karoji}, {Furusawa}, {Kawanomoto}, {Morokuma}, {Ishizuka},
  {Nariai}, {Tanaka}, {Uraguchi}, {Utsumi}, {Obuchi}, {Okura}, {Oguri},
  {Takata}, {Tomono}, {Kurakami}, {Namikawa}, {Usuda}, {Yamanoi}, {Terai},
  {Uekiyo}, {Yamada}, {Koike}, {Aihara}, {Fujimori}, {Mineo}, {Miyatake},
  {Yasuda}, {Nishizawa}, {Saito}, {Tanaka}, {Uchida}, {Katayama}, {Wang},
  {Chen}, {Lupton}, {Loomis}, {Bickerton}, {Price}, {Gunn}, {Suzuki},
  {Miyazaki}, {Muramatsu}, {Yamamoto}, {Endo}, {Ezaki}, {Itoh}, {Miwa},
  {Yokota}, {Matsuda}, {Ebinuma}, \& {Takeshi}}]{Miyazaki_hsc}
{Miyazaki}, S., {Komiyama}, Y., {Nakaya}, H., {et~al.} 2012, in Society of
  Photo-Optical Instrumentation Engineers (SPIE) Conference Series, Vol. 8446,
  Ground-based and Airborne Instrumentation for Astronomy IV, 84460Z

\bibitem[{{More} {et~al.}(2016{\natexlab{a}}){More}, {Oguri}, {Kayo}, {Zinn},
  {Strauss}, {Santiago}, {Mosquera}, {Inada}, {Kochanek}, {Rusu}, {Brownstein},
  {da Costa}, {Kneib}, {Maia}, {Quimby}, {Schneider}, {Streblyanska}, \&
  {York}}]{BQLS}
{More}, A., {Oguri}, M., {Kayo}, I., {et~al.} 2016{\natexlab{a}}, \mnras, 456,
  1595

\bibitem[{{More} {et~al.}(2016{\natexlab{b}}){More}, {Verma}, {Marshall},
  {More}, {Baeten}, {Wilcox}, {Macmillan}, {Cornen}, {Kapadia}, {Parrish},
  {Snyder}, {Davis}, {Gavazzi}, {Lintott}, {Simpson}, {Miller}, {Smith},
  {Paget}, {Saha}, {K{\"u}ng}, \& {Collett}}]{space_warp_02}
{More}, A., {Verma}, A., {Marshall}, P.~J., {et~al.} 2016{\natexlab{b}},
  \mnras, 455, 1191

\bibitem[{{Myers} {et~al.}(2003){Myers}, {Jackson}, {Browne}, {de Bruyn},
  {Pearson}, {Readhead}, {Wilkinson}, {Biggs}, {Blandford}, {Fassnacht},
  {Koopmans}, {Marlow}, {McKean}, {Norbury}, {Phillips}, {Rusin}, {Shepherd},
  \& {Sykes}}]{class_01}
{Myers}, S.~T., {Jackson}, N.~J., {Browne}, I.~W.~A., {et~al.} 2003, \mnras,
  341, 1

\bibitem[{{Nierenberg} {et~al.}(2020){Nierenberg}, {Gilman}, {Treu}, {Brammer},
  {Birrer}, {Moustakas}, {Agnello}, {Anguita}, {Fassnacht}, {Motta}, {Peter},
  \& {Sluse}}]{anna_dm_sub}
{Nierenberg}, A.~M., {Gilman}, D., {Treu}, T., {et~al.} 2020, \mnras, 492, 5314

\bibitem[{{Oguri}(2010)}]{Masamune_glafic}
{Oguri}, M. 2010, \pasj, 62, 1017

\bibitem[{{Oguri} {et~al.}(2006){Oguri}, {Inada}, {Pindor}, {Strauss},
  {Richards}, {Hennawi}, {Turner}, {Lupton}, {Schneider}, {Fukugita}, \&
  {Brinkmann}}]{sqls_01}
{Oguri}, M., {Inada}, N., {Pindor}, B., {et~al.} 2006, \aj, 132, 999

\bibitem[{{Oguri} \& {Marshall}(2010)}]{OM10}
{Oguri}, M. \& {Marshall}, P.~J. 2010, \mnras, 405, 2579

\bibitem[{{Ostrovski} {et~al.}(2018){Ostrovski}, {Lemon}, {Auger}, {McMahon},
  {Fassnacht}, {Chen}, {Connolly}, {Koposov}, {Pons}, {Reed}, \&
  {Rusu}}]{Ostrovski_gaia_panstarrs_03}
{Ostrovski}, F., {Lemon}, C.~A., {Auger}, M.~W., {et~al.} 2018, \mnras, 473,
  L116

\bibitem[{{Ostrovski} {et~al.}(2017){Ostrovski}, {McMahon}, {Connolly},
  {Lemon}, {Auger}, {Banerji}, {Hung}, {Koposov}, {Lidman}, {Reed}, {Allam},
  {Benoit-L{\'e}vy}, {Bertin}, {Brooks}, {Buckley-Geer}, {Carnero Rosell},
  {Carrasco Kind}, {Carretero}, {Cunha}, {da Costa}, {Desai}, {Diehl},
  {Dietrich}, {Evrard}, {Finley}, {Flaugher}, {Fosalba}, {Frieman}, {Gerdes},
  {Goldstein}, {Gruen}, {Gruendl}, {Gutierrez}, {Honscheid}, {James}, {Kuehn},
  {Kuropatkin}, {Lima}, {Lin}, {Maia}, {Marshall}, {Martini}, {Melchior},
  {Miquel}, {Ogando}, {Plazas Malag{\'o}n}, {Reil}, {Romer}, {Sanchez},
  {Santiago}, {Scarpine}, {Sevilla-Noarbe}, {Soares-Santos}, {Sobreira},
  {Suchyta}, {Tarle}, {Thomas}, {Tucker}, \& {Walker}}]{Strides_method}
{Ostrovski}, F., {McMahon}, R.~G., {Connolly}, A.~J., {et~al.} 2017, \mnras,
  465, 4325

\bibitem[{{Peng} {et~al.}(2006){Peng}, {Impey}, {Rix}, {Kochanek}, {Keeton},
  {Falco}, {Leh{\'a}r}, \& {McLeod}}]{quasar_host_01}
{Peng}, C.~Y., {Impey}, C.~D., {Rix}, H.-W., {et~al.} 2006, \apj, 649, 616

\bibitem[{{Rusu} {et~al.}(2019){Rusu}, {Berghea}, {Fassnacht}, {More}, {Seman},
  {Nelson}, \& {Chen}}]{Rusu_18}
{Rusu}, C.~E., {Berghea}, C.~T., {Fassnacht}, C.~D., {et~al.} 2019, \mnras,
  486, 4987

\bibitem[{{S{\'a}nchez} \& {Des Collaboration}(2010)}]{des_survey}
{S{\'a}nchez}, E. \& {Des Collaboration}. 2010, in Journal of Physics
  Conference Series, Vol. 259, Journal of Physics Conference Series, 012080

\bibitem[{{Scoville} {et~al.}(2007){Scoville}, {Aussel}, {Brusa}, {Capak},
  {Carollo}, {Elvis}, {Giavalisco}, {Guzzo}, {Hasinger}, {Impey}, {Kneib},
  {LeFevre}, {Lilly}, {Mobasher}, {Renzini}, {Rich}, {Sanders}, {Schinnerer},
  {Schminovich}, {Shopbell}, {Taniguchi}, \& {Tyson}}]{COSMOS}
{Scoville}, N., {Aussel}, H., {Brusa}, M., {et~al.} 2007, \apjs, 172, 1

\bibitem[{{Silverman} {et~al.}(2011){Silverman}, {Kampczyk}, {Jahnke},
  {Andrae}, {Lilly}, {Elvis}, {Civano}, {Mainieri}, {Vignali}, {Zamorani},
  {Nair}, {Le F{\`e}vre}, {de Ravel}, {Bardelli}, {Bongiorno}, {Bolzonella},
  {Cappi}, {Caputi}, {Carollo}, {Contini}, {Coppa}, {Cucciati}, {de la Torre},
  {Franzetti}, {Garilli}, {Halliday}, {Hasinger}, {Iovino}, {Knobel},
  {Koekemoer}, {Kova{\v{c}}}, {Lamareille}, {Le Borgne}, {Le Brun}, {Maier},
  {Mignoli}, {Pello}, {P{\'e}rez-Montero}, {Ricciardelli}, {Peng}, {Scodeggio},
  {Tanaka}, {Tasca}, {Tresse}, {Vergani}, {Zucca}, {Brusa}, {Cappelluti},
  {Comastri}, {Finoguenov}, {Fu}, {Gilli}, {Hao}, {Ho}, \&
  {Salvato}}]{silverman_merger}
{Silverman}, J.~D., {Kampczyk}, P., {Jahnke}, K., {et~al.} 2011, \apj, 743, 2

\bibitem[{{Sonnenfeld} {et~al.}(2018){Sonnenfeld}, {Chan}, {Shu}, {More},
  {Oguri}, {Suyu}, {Wong}, {Lee}, {Coupon}, {Yonehara}, {Bolton}, {Jaelani},
  {Tanaka}, {Miyazaki}, \& {Komiyama}}]{sugohi}
{Sonnenfeld}, A., {Chan}, J. H.~H., {Shu}, Y., {et~al.} 2018, \pasj, 70, S29

\bibitem[{{Sonnenfeld} {et~al.}(2020){Sonnenfeld}, {Verma}, {More}, {Baeten},
  {Macmillan}, {Wong}, {Chan}, {Jaelani}, {Lee}, {Oguri}, {Rusu}, {Veldthuis},
  {Trouille}, {Marshall}, {Hutchings}, {Allen}, {O'Donnell}, {Cornen}, {Davis},
  {McMaster}, {Lintott}, \& {Miller}}]{ale_citizen}
{Sonnenfeld}, A., {Verma}, A., {More}, A., {et~al.} 2020, \aap, 642, A148

\bibitem[{{Spiniello} {et~al.}(2018){Spiniello}, {Agnello}, {Napolitano},
  {Sergeyev}, {Getman}, {Tortora}, {Spavone}, {Bilicki}, {Buddelmeijer},
  {Koopmans}, {Kuijken}, {Vernardos}, {Bannikova}, \&
  {Capaccioli}}]{Spinello_gaia_kids}
{Spiniello}, C., {Agnello}, A., {Napolitano}, N.~R., {et~al.} 2018, \mnras,
  480, 1163

\bibitem[{{Stemo} {et~al.}(2020){Stemo}, {Comerford}, {Barrows}, {Stern},
  {Assef}, {Griffith}, \& {Schechter}}]{stemo_merger}
{Stemo}, A., {Comerford}, J.~M., {Barrows}, R.~S., {et~al.} 2020, arXiv
  e-prints, arXiv:2011.10051

\bibitem[{{Williams} {et~al.}(2017){Williams}, {Agnello}, \&
  {Treu}}]{Peter_Williams_method}
{Williams}, P., {Agnello}, A., \& {Treu}, T. 2017, \mnras, 466, 3088

\bibitem[{{Wong} {et~al.}(2018){Wong}, {Sonnenfeld}, {Chan}, {Rusu}, {Tanaka},
  {Jaelani}, {Lee}, {More}, {Oguri}, {Suyu}, \& {Komiyama}}]{ken_sugohi}
{Wong}, K.~C., {Sonnenfeld}, A., {Chan}, J. H.~H., {et~al.} 2018, \apj, 867,
  107

\bibitem[{{Wong} {et~al.}(2020){Wong}, {Suyu}, {Chen}, {Rusu}, {Millon},
  {Sluse}, {Bonvin}, {Fassnacht}, {Taubenberger}, {Auger}, {Birrer}, {Chan},
  {Courbin}, {Hilbert}, {Tihhonova}, {Treu}, {Agnello}, {Ding}, {Jee},
  {Komatsu}, {Shajib}, {Sonnenfeld}, {Blandford}, {Koopmans}, {Marshall}, \&
  {Meylan}}]{holicow_h0_01}
{Wong}, K.~C., {Suyu}, S.~H., {Chen}, G. C.~F., {et~al.} 2020, \mnras, 498,
  1420

\bibitem[{{Yasuda} {et~al.}(2019){Yasuda}, {Tanaka}, {Tominaga}, {Jiang},
  {Moriya}, {Morokuma}, {Suzuki}, {Takahashi}, {Yamaguchi}, {Maeda}, {Sako},
  {Ikeda}, {Kimura}, {Morii}, {Ueda}, {Yoshida}, {Lee}, {Suyu}, {Komiyama},
  {Regnault}, \& {Rubin}}]{hsc_transient}
{Yasuda}, N., {Tanaka}, M., {Tominaga}, N., {et~al.} 2019, \pasj, 71, 74

\bibitem[{{York} {et~al.}(2000){York}, {Adelman}, {Anderson}, {Anderson},
  {Annis}, {Bahcall}, {Bakken}, {Barkhouser}, {Bastian}, {Berman}, {Boroski},
  {Bracker}, {Briegel}, {Briggs}, {Brinkmann}, {Brunner}, {Burles}, {Carey},
  {Carr}, {Castander}, {Chen}, {Colestock}, {Connolly}, {Crocker}, {Csabai},
  {Czarapata}, {Davis}, {Doi}, {Dombeck}, {Eisenstein}, {Ellman}, {Elms},
  {Evans}, {Fan}, {Federwitz}, {Fiscelli}, {Friedman}, {Frieman}, {Fukugita},
  {Gillespie}, {Gunn}, {Gurbani}, {de Haas}, {Haldeman}, {Harris}, {Hayes},
  {Heckman}, {Hennessy}, {Hindsley}, {Holm}, {Holmgren}, {Huang}, {Hull},
  {Husby}, {Ichikawa}, {Ichikawa}, {Ivezi{\'c}}, {Kent}, {Kim}, {Kinney},
  {Klaene}, {Kleinman}, {Kleinman}, {Knapp}, {Korienek}, {Kron}, {Kunszt},
  {Lamb}, {Lee}, {Leger}, {Limmongkol}, {Lindenmeyer}, {Long}, {Loomis},
  {Loveday}, {Lucinio}, {Lupton}, {MacKinnon}, {Mannery}, {Mantsch}, {Margon},
  {McGehee}, {McKay}, {Meiksin}, {Merelli}, {Monet}, {Munn}, {Narayanan},
  {Nash}, {Neilsen}, {Neswold}, {Newberg}, {Nichol}, {Nicinski}, {Nonino},
  {Okada}, {Okamura}, {Ostriker}, {Owen}, {Pauls}, {Peoples}, {Peterson},
  {Petravick}, {Pier}, {Pope}, {Pordes}, {Prosapio}, {Rechenmacher}, {Quinn},
  {Richards}, {Richmond}, {Rivetta}, {Rockosi}, {Ruthmansdorfer}, {Sandford},
  {Schlegel}, {Schneider}, {Sekiguchi}, {Sergey}, {Shimasaku}, {Siegmund},
  {Smee}, {Smith}, {Snedden}, {Stone}, {Stoughton}, {Strauss}, {Stubbs},
  {SubbaRao}, {Szalay}, {Szapudi}, {Szokoly}, {Thakar}, {Tremonti}, {Tucker},
  {Uomoto}, {Vanden Berk}, {Vogeley}, {Waddell}, {Wang}, {Watanabe},
  {Weinberg}, {Yanny}, {Yasuda}, \& {SDSS Collaboration}}]{SDSS_survey}
{York}, D.~G., {Adelman}, J., {Anderson}, Jr., J.~E., {et~al.} 2000, \aj, 120,
  1579

\end{thebibliography}

\end{document}